\documentclass[11pt]{article} 
\usepackage{amssymb,amsmath,latexsym,bm,amsfonts,bbm}

\usepackage[left=3cm, right=3cm, top=3cm]{geometry}

\RequirePackage[T1]{fontenc}

\RequirePackage{graphicx}
\RequirePackage{mathptmx}      
\RequirePackage{flushend}
\RequirePackage[numbers,sort&compress]{natbib}
\RequirePackage[colorlinks,citecolor=blue,urlcolor=blue,linkcolor=blue]{hyperref}

\def\simlt{\mathrel{\lower2.5pt\vbox{\lineskip=0pt\baselineskip=0pt
           \hbox{$<$}\hbox{$\sim$}}}}
\def\simgt{\mathrel{\lower2.5pt\vbox{\lineskip=0pt\baselineskip=0pt
           \hbox{$>$}\hbox{$\sim$}}}}

\begin{document}

\begin{titlepage}
  \ \\
  
  \begin{center}
    \vspace{1 cm}

\huge{\bf Inflation near a metastable de Sitter vacuum from moduli stabilisation}

    \vspace{1.8 cm}
    
\large{Ignatios Antoniadis\footnote{e-mail: antoniad@lpthe.jussieu.fr}$^{a,b}$,  \, \, Osmin Lacombe\footnote{e-mail: osmin@lpthe.jussieu.fr}$^{a}$, \, \, George K. Leontaris\footnote{{e-mail: leonta@uoi.gr}}$^{c}$
}

    \vspace{1.5cm}

\normalsize{\it $^a$ Laboratoire de Physique Th\'eorique et Hautes Energies - LPTHE, Sorbonne Universit\'e, CNRS, \\4 Place Jussieu, 75005 Paris, France 

$^b$ Albert Einstein Center, Institute for Theoretical Physics, University of Bern,  \\ Sidlerstrasse 5, CH-3012 Bern, Switzerland 

$^c$ Physics Department, University of Ioannina, 45110, Ioannina, Greece}

    \vspace{1.5cm}

\begin{abstract}
We study the cosmological properties of a metastable de Sitter vacuum obtained recently in the framework of type IIB flux compactifications in the presence of three D7-brane stacks, based on perturbative quantum corrections at both world-sheet and string loop level that are dominant at large volume and weak coupling. In the simplest case, the model has one effective parameter controlling the shape of the potential of the inflaton which is identified with the volume modulus. The model provides a phenomenological successful small-field inflation for a value of the parameter that makes the minimum very shallow and near the maximum. The horizon exit is close to the inflection point while most of the required e-folds of the Universe expansion are generated near the minimum, with a prediction for the ratio of tensor-to-scalar primordial fluctuations $r\simeq 4\times 10^{-4}$. Despite its shallowness, the minimum turns out to be practically stable. We show that it can decay only through the Hawking-Moss instanton leading to an extremely long decay rate. Obviously, in order to end inflation and obtain a realistic model, new low-energy physics is needed around the minimum, at intermediate energy scales of order $10^{12}$  GeV. An attractive possibility is by introducing a ``waterfall'' field within the framework of hybrid inflation.

\end{abstract}

\end{center}

\end{titlepage}

\newpage
\pagestyle{plain}
\pagenumbering{arabic}

 {\hypersetup{linkcolor=black}
 \tableofcontents
}

\newpage

\section{Introduction}

	Inflation has been the most prevalent  scenario for solving   the horizon and flatness problems~\cite{Linde:1981mu,Liddle:2000cg}. It is usually realised using a scalar field	coupled to gravity, which 
	rolls down towards a (local) minimum of the potential (which is not necessarily the true minimum). In the ensuing years several elaborated versions appeared, where more than
	one fields are included to drive inflation. For example, in the
	case of the hybrid scenario~\cite{Linde:1993cn} inflation ends when a second `waterfall' scalar field takes over and rolls	rapidly down to the slope 	towards  the true minimum that should describe our observable Universe.  In many  effective theories with ultraviolet (UV) completion
	this  is a common way to implement the inflationary scenario,  as the   scalar potential of 	the theory involves several scalar fields defining  trajectories with metastable minima before a true minimum 
	has entailed.
	
	 In a wide class of  string theory  constructions, the inflaton field is materialised  by some modulus, since
	 there are  always  plenty of them in generic compactifications, parametrising in particular the internal manifold.  
  However, an important step for
  a successful implementation of the inflationary scenario
  in string theory is the existence of  a (metastable) de Sitter vacuum
  of the corresponding effective field theory \cite{Kachru:2003aw,Conlon:2005ki}.
  Furthermore, in the broader picture the viability of such a construction is 
   intimately related to the  important issue of moduli stabilisation, 
   providing positive square-masses for all the associated scalar fields.

In the present work, we  investigate  the issue of cosmological inflation in a class of effective models emerging in the framework
of type-IIB string theory and its extended geometric picture, namely the F-theory. 
Within this context, in Ref~\cite{Antoniadis:2018hqy, Antoniadis:2019rkh} it was shown that when 
the internal geometry is equipped with an appropriate configuration
of  intersecting D7-branes and orbifold O7-planes, 
  string loop effects induce logarithmic corrections due to closed strings propagating effectively along the two-dimensional 
  space~\cite{Antoniadis:1998ax} transverse to  each D7-brane towards points of localised gravity~\cite{Antoniadis:2002tr}. Working in the large volume limit, 
 it has been demonstrated how the associated K\"ahler moduli 
  can be stabilised and an uplift of the vacuum energy can be ensured by  D-terms
 related to anomalous $U(1)$ symmetries on the D7-branes world-volume \cite{Burgess:2003ic}.

A first attempt to implement inflation in this framework using the volume modulus as inflaton was done in~\cite{Antoniadis:2018ngr}, in the presence of a new Fayet-Iliopoulos term written recently in supergravity~\cite{Cribiori:2017laj,Antoniadis:2018oeh} which provides an extra uplift of the vacuum energy and allows a tuning of its value at the minimum. The origin of such a term is however unclear in string theory making questionable the applicability of the stabilisation mechanism to inflation.
In the present work, instead, we follow an alternative path, focusing only on the minimal amount of ingredients which were already used to  ensure moduli stabilisation at a metastable de Sitter vacuum. Our first goal is to investigate the possibility of realistic inflation without imposing the present tiny value of the vacuum energy at the minimum of the scalar potential. We will then address this issue together with the question of ending inflation by introducing a ``waterfall'' field in the context of hybrid inflation.

We recall first briefly the main ingredients of the stabilisation mechanism. 
All complex structure moduli and the axion-dilaton field are fixed in a supersymmetric way
(imposing vanishing F-terms) on the fluxed induced superpotential~\cite{Frey:2002hf,Kachru:2002he}.
 We end up in general with a constant superpotential but vanishing scalar potential for the K\"ahler moduli due to the no-scale structure of their kinetic terms at the leading order~\cite{Cremmer:1983bf,Ellis:1983ei,Becker:2002nn}.
Radiative corrections though break the no-scale structure and bring a logarithmic dependence on the co-dimension two volumes transverse to the D7-branes world-volumes~\cite{Antoniadis:2018hqy,Antoniadis:2019rkh}.
Moreover, D-term contributions from anomalous $U(1)$ factors associated with the
 intersecting D7-branes provide an uplift mechanism  \cite{Burgess:2003ic} so that 
 a de Sitter (dS) vacuum is naturally achieved.

 In the simplest case,
 the scalar potential of the effective four dimensional  theory can 
 be expressed in terms of the volume modulus and two other
 (orthogonal combinations)  K\"ahler moduli fields. 
 It turns out that essentially only one free parameter, dubbed $x$ in the following, controls the shape of the 
 potential and in particular delimits its 
 two extrema (minimum and maximum). 
More precisely, the requirement of a dS minimum
confines   $x$ in a very small region where the potential stays almost flat, and  the two
extrema of the potential are very close to each other. 
In effect, $x$   regulates the measurable 
parameters related to inflation. 
 
 In this restrictive context, we start our investigations by  examining  in some detail possible implementations of various existing  inflationary scenarios including slow roll and in particular the hilltop inflation. Varying $x$, we adjust the value
 of the slow-roll parameter $\eta$ so that inflation starts near the maximum with the correct value of the spectral index. However in this case the slow-roll parameters remain small all the way to the minimum and inflation doesn't stop, producing much more than the required 60 e-folds. Alternatively, imposing the correct number of e-folds, the resulting spectral index does not reproduce the observable value. Thus, the hilltop possibility is ruled out.

We then proceed with a novel  proposal where the horizon exit occurs near the inflection point of the potential where the $\eta$ parameter vanishes and inflation takes place essentially near the minimum where the required number of e-folds are produced. 
This is  a reasonable and completely justifiable assumption at the other end,
 however, in close analogy with the
 concept of hilltop inflation. Indeed, both are characterised with 
 the property that the slow-roll parameter $\epsilon$ is  negligible 
 at  both extrema and consequently, a short interval of the inflaton
 trajectory   is enough to accumulate the required number of e-folds. 
Remarkably, we find that an inflationary phase is feasible 
near the minimum and the desired number of e-folds can readily be achieved. 
It also predicts a ratio of tensor-to-scalar primordial fluctuations $r\simeq 4\times 10^{-4}$.
Moreover, because of the proximity of the two extrema of the potential, the inflaton is restricted to a short range of values ensuring small field inflation, compatible with the validity of the effective field theory.

On the other hand, since the minimum is generated from quantum corrections, it is metastable and
is expected to decay to the true minimum in the runaway direction of  large volume.
We then perform an estimate of its lifetime due to either tunnelling by the Coleman-de Lucia 
instanton~\cite{Coleman:1977py, Coleman:1980aw}, or passing over the barrier by the Hawking-Moss instanton~\cite{Hawking:1981fz}. Our analysis shows that in the $x$-region where inflation 
is viable the false vacuum decay is due to the latter, leading to an extremely long lifetime. 
 
 The paper is organised as follows. In Section 2, we present a short review of the framework and the main features of the mechanism of moduli stabilisation (subsection 2.1), as well as of the minimisation of the potential that depends on one effective parameter. In Section 3, we start with the basic equations for inflationary solutions (subsection 3.1) and perform the analysis for hilltop inflation (subsection 3.2) and inflation around the minimum from the inflection point (subsection 3.3). We then work out the observable quantities and compare the parameters of the model with those of the underlying string theory (subsection 3.4). In Section 4, we compute the vacuum decay (subsection 4.1) and discuss the issue of ending inflation in the context of the hybrid proposal by introducing a ``waterfall'' field (subsection 4.2). We conclude in Section 5 with a summary of the main results.

\section{Scalar potential from D7-branes moduli stabilisation} \label{themodel}
\subsection{Type IIB model of intersecting D7-branes and moduli stabilisation}
In this paper we consider the model developed in \cite{Antoniadis:2018hqy} within the type IIB string framework where complex structure moduli and the dilaton are stabilised in the standard supersymmetric way by turning on 3-form fluxes. This model takes into account the quantum corrections of a three intersecting D7-branes configuration \cite{Antoniadis:2019rkh}. These corrections break the no-scale structure of the effective theory and give a non-zero contribution to the F-part of the supergravity scalar potential. If one also considers the $U(1)$ anomalous symmetries of the D7-branes, Fayet-Iliopoulos D-terms must be introduced in the scalar potential \cite{Burgess:2003ic}. The latter can be used to uplift the scalar potential to a de Sitter minimum with all K\"ahler moduli stabilised. Denoting by $\tau_i$ for $i=1,2,3$ the imaginary parts of the D7-branes world-volume K\"ahler moduli,  the K\"ahler potential of the model is \cite{Antoniadis:2018hqy} \footnote{For simplicity, we drop off the real parts which are absorbed by the anomalous $U(1)$'s to become massive, and play no role in the minimisation procedure.}
\begin{equation}
\mathcal{K}=-\frac{2}{\kappa^2}\ln\left((\tau_1\tau_2\tau_3)^{\frac 12}+\xi+\sum_{i=1}^{3}\gamma_{i}\ln(\tau_i)\right),\label{Kahler}
\end{equation}
where  $\kappa=\sqrt{8\pi G}$ is the inverse of the reduced Planck mass (we work with $c=\hbar=1$) and the compactification volume $\mathcal{V}$ is expressed (in Planck units) as
\begin{equation}
\mathcal{V}=(\tau_1\tau_2\tau_3)^{\frac 12}.
\end{equation}

The $\xi$ term stands for $\alpha'^3$ corrections \cite{Becker:2002nn} and is proportional to the Euler number  $\chi_{CY}$ of the Calabi-Yau manifold
\begin{equation}
\xi= -\frac{\zeta(3)}{4} \chi_{CY}\,. 
\label{xi}
\end{equation}
In the large volume limit, it induces four-dimensional kinetic terms~\cite{Antoniadis:2002tr}.
In the case of orbifolds, it is generated at one string loop order and reads: $\xi_{orb}=-{\pi^2\chi_{orb}g_s^2/12}$, where $g_s$ is the string coupling and $\chi_{orb}$ counts the difference between the numbers of closed string $\mathcal{N}=2$ hyper and vector multiplets, $\chi_{orb}=4(n_H-n_V)$.
The $\gamma_i$ are model dependant parameters for the logarithmic quantum corrections associated with the 7-branes \cite{Antoniadis:2019rkh}. They are induced from massless closed strings emitted from the localised vertices towards the 7-brane sources, thus propagating in two dimensions. Taking identical $\gamma_i$ for simplicity, they are given by~\footnote{Here we assumed that the total Euler number is concentrated at the same point in the large volume limit; in general $\xi$ in \eqref{gammas} is part of $\xi$ in \eqref{xi}.}
\begin{equation}
 \gamma_1=\gamma_2=\gamma_3 \equiv \gamma = -\frac 12 g_s T_0 \, \xi 
 \label{gammas},
\end{equation}
where $T_0/g_s$ is the effective 7-brane tension. Note the minus sign in the last equality of \eqref{gammas}. 

One can extract the F-part of the scalar potential from \eqref{Kahler}. It  depends only on the volume $\mathcal{V}$ and after defining $\mu=e^{\frac{\xi}{2\gamma}}$, its exact expression is
\begin{equation}
V_F=\-\frac{3\gamma \mathcal{W}_0^2}{\kappa^4}\frac{2(\gamma+2\mathcal{V})+(4\gamma-\mathcal{V})\ln(\mu\mathcal{V})}{\left(\mathcal{V}+2\gamma\ln(\mu\mathcal{V})\right)^2\left(6\gamma^2+\mathcal{V}^2+8\gamma\mathcal{V}+\gamma(4\gamma-\mathcal{V})\ln(\mu\mathcal{V})\right)},
\end{equation}
where $\mathcal{W}_0$ is the constant superpotential contribution left over from the 3-form fluxes upon the complex structure moduli stabilisation.
In the large volume limit, $V_F$ takes the much simpler form
\begin{equation}
V_F= \frac{3 \mathcal{W}_0^2}{2\kappa^4 \mathcal{V}^3}\left(2\gamma(\ln \mathcal{V}-4)+\xi \right) + \cdots \label{LVL}
\end{equation}
 
The D-part of the scalar potential coming from the D7-branes can also be expressed very simply in the large world-volume limit
\begin{align}
V_D=\frac{d_1}{\kappa^4\tau_1^3}+\frac{d_2}{\kappa^4\tau_2^3}+\frac{d_3}{\kappa^4\tau_3^3}+ \cdots \label{VD}
\end{align}
where the $d_i$ for $i=1,2,3$ are model dependent constants related to the $U(1)$ anomalies. Contrary to the F-part, this D-part depends on the three $\tau_i$ fields. 
Instead of these K\"ahler moduli we will rather work with the canonically normalised fields
\begin{align}
t_i=\frac 1{\sqrt{2}} \ln(\tau_i),
\end{align}
from which we obtain the following base, after isolating the volume from the two other perpendicular directions
\begin{align}
&\phi=\frac{1}{\sqrt{3}}(t_1+t_2+t_3)=\frac{\sqrt{6}}{3}\ln(\mathcal{V}), & \label{tandV}\\
&u=\frac{1}{\sqrt{2}}(t_1-t_2),&\\
&v=\frac{1}{\sqrt{6}}(t_1+t_2-2t_3).&
\end{align}
In terms of these fields, the D-part of the potential \eqref{VD} reads
\begin{equation}
V_D=\frac{e^{-\sqrt{6}\phi}}{\kappa^4}\left(d_1e^{-\sqrt{3}v-3u}+d_2e^{-\sqrt{3}v+3u}+d_3e^{2\sqrt{3}v}\right)+\cdots \label{VDcanonical}
\end{equation}
so that the total scalar potential is
\begin{equation}
V_F+V_D=\frac{3 \mathcal{W}_0^2}{2\kappa^4}e^{-3 \sqrt{\frac 32} \phi} \left(\gamma \left(\sqrt{6}\phi-4\right)+\xi \right)+\frac{e^{-\sqrt{6}\phi}}{\kappa^4}\left(d_1e^{-\sqrt{3}v-3u}+d_2e^{-\sqrt{3}v+3u}+d_3e^{2\sqrt{3}v}\right)+\cdots  \label{Vtot}
\end{equation}

\subsection{Local de Sitter minimum}\label{dSminimumpot}

We study now the minimum of the scalar potential \eqref{Vtot}. The field $\phi$ will be associated with the inflaton and its evolution will determine the inflation era. We must then stabilise the two other canonically normalised fields $u$, $v$ at their values $u_0$ and $v_0$ dictated by the minimisation of $V_D$ in \eqref{VD}. Their values at the minimum read 
\begin{equation}
u_0=\frac 16 \ln\left(\frac{d_1}{d_2}\right), \quad v_0=\frac 1{6\sqrt{3}} \ln\left(\frac{d_1d_2}{d_3^2}\right),
\end{equation}
for which the potential $V_D$ becomes
\begin{equation}
V_D(\phi,u_0,v_0)=\frac{3(d_1d_2d_3)^{\frac 13}}{\kappa^4\mathcal{V}^2}=\frac d{\kappa^4\mathcal{V}^2}=\frac{d}{\kappa^4} e^{-\sqrt{6}\phi},
\end{equation}
with $d\equiv3(d_1d_2d_3)^{\frac 13}$. Hence after stabilisation of the two transverse moduli, the total scalar potential reduces in the large volume limit to
%
\begin{align}
V (\mathcal{V})=V_F+V_D&\simeq \frac{3 \mathcal{W}_0^2}{2\kappa^4 \mathcal{V}^3}\left(2\gamma(\ln \mathcal{V}-4)+\xi \right) +\frac d{\kappa^4\mathcal{V}^2}\equiv \frac{C}{\kappa^4} \left( -\frac{\ln \mathcal{V}-4+q}{{\mathcal V}^3}-\frac{3\sigma}{2\mathcal{V}^2}\right), \label{Vlargelimit}
\end{align}
where we defined 
\begin{equation}
q\equiv \frac \xi{2 \gamma}, \quad \sigma \equiv \frac{2d}{9 {\mathcal{W}_0}^2\gamma}, \quad C\equiv -3{\mathcal{W}_0}^2\gamma>0. \label{modelparameters}
\end{equation}
The last inequality is obtained for $\gamma<0$, which is a condition for a dS vacuum to exist at large volume. The parameter $q$ essentially shifts the local extrema towards large volumes. It is essential in the string context but does not play a role for inflation. Thus, for simplicity, we will take it zero in the numerical study of section \ref{sectioninflation},  before coming back to its significance in section \ref{physicalparameters}. $C$ is an overall constant which plays no role in the minimisation but will be related to the observed spectral amplitude, when the model is considered as a candidate for an inflationary scenario. Thus, $\sigma$ is the only effective parameter of the model.

\begin{figure}[h] 
\makebox[\textwidth]{\includegraphics[scale=0.55]{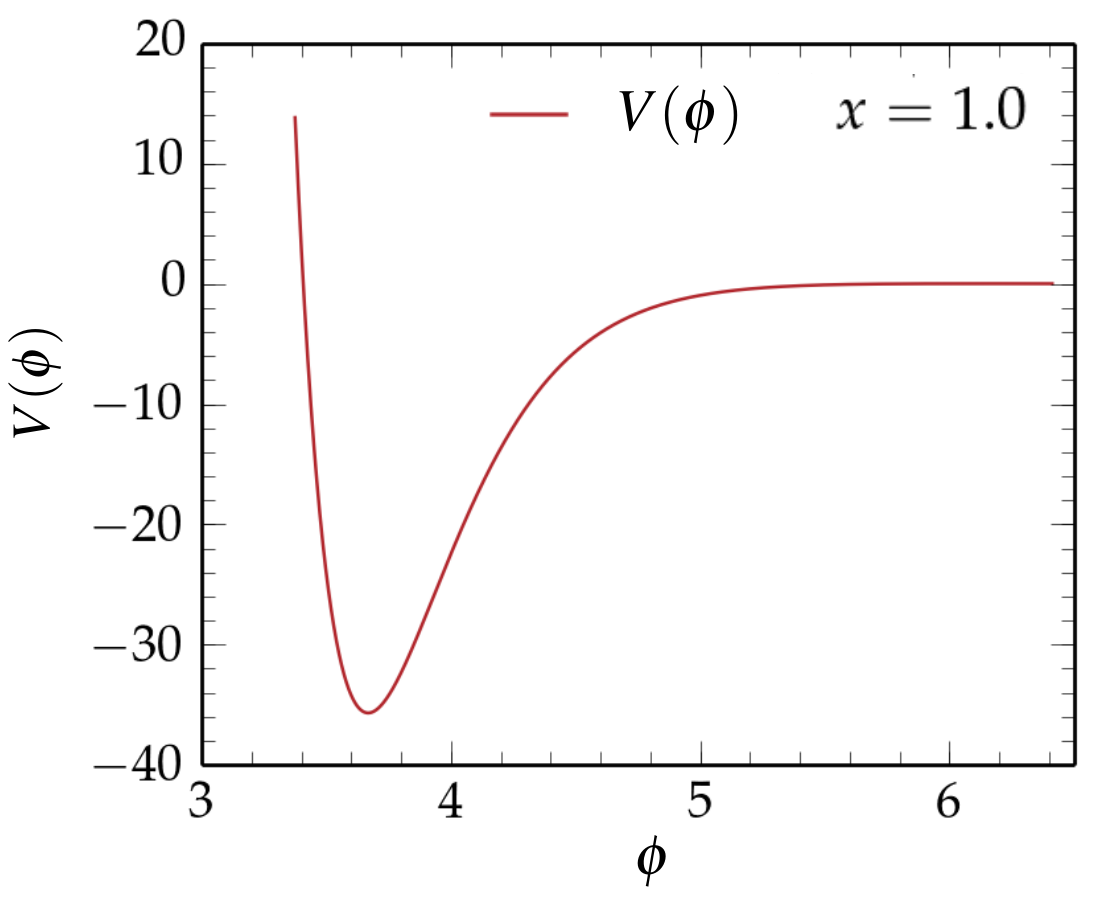}\vspace{-0.0cm}
 \includegraphics[scale=0.55]{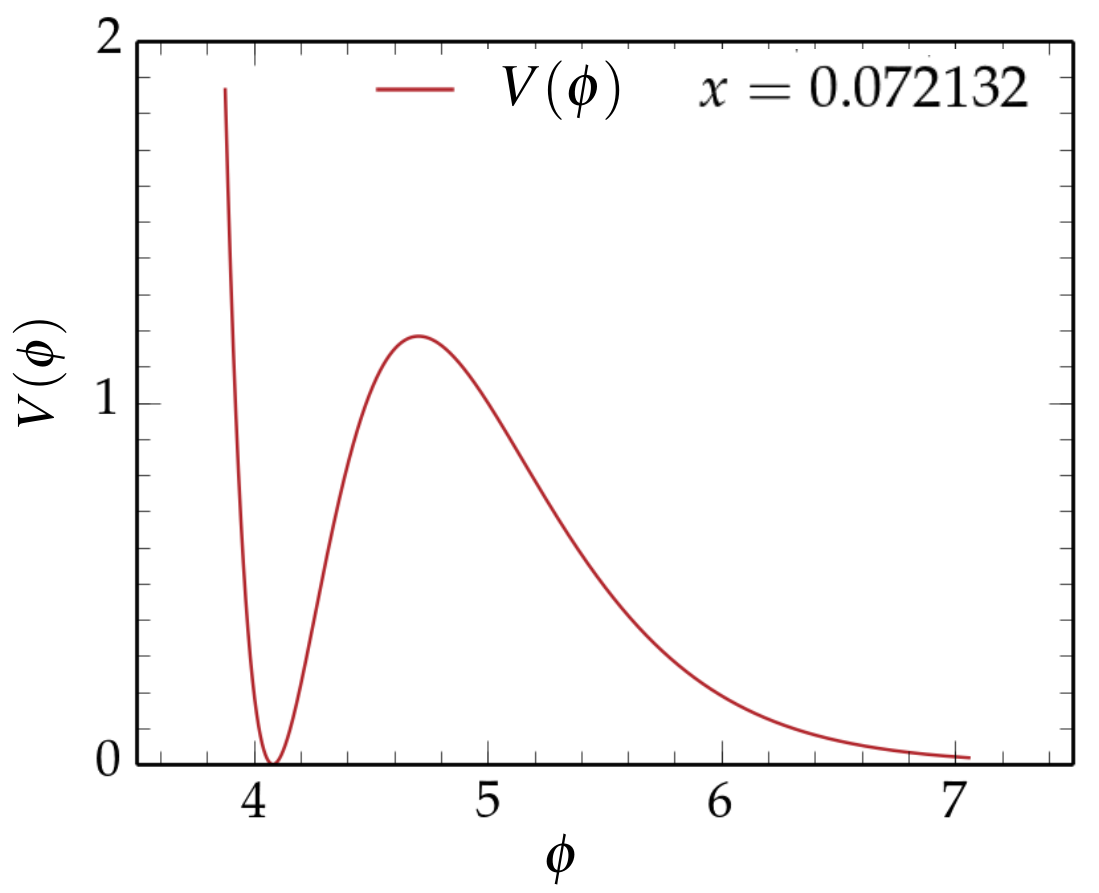}\vspace{-0.3cm}
\includegraphics[scale=0.55]{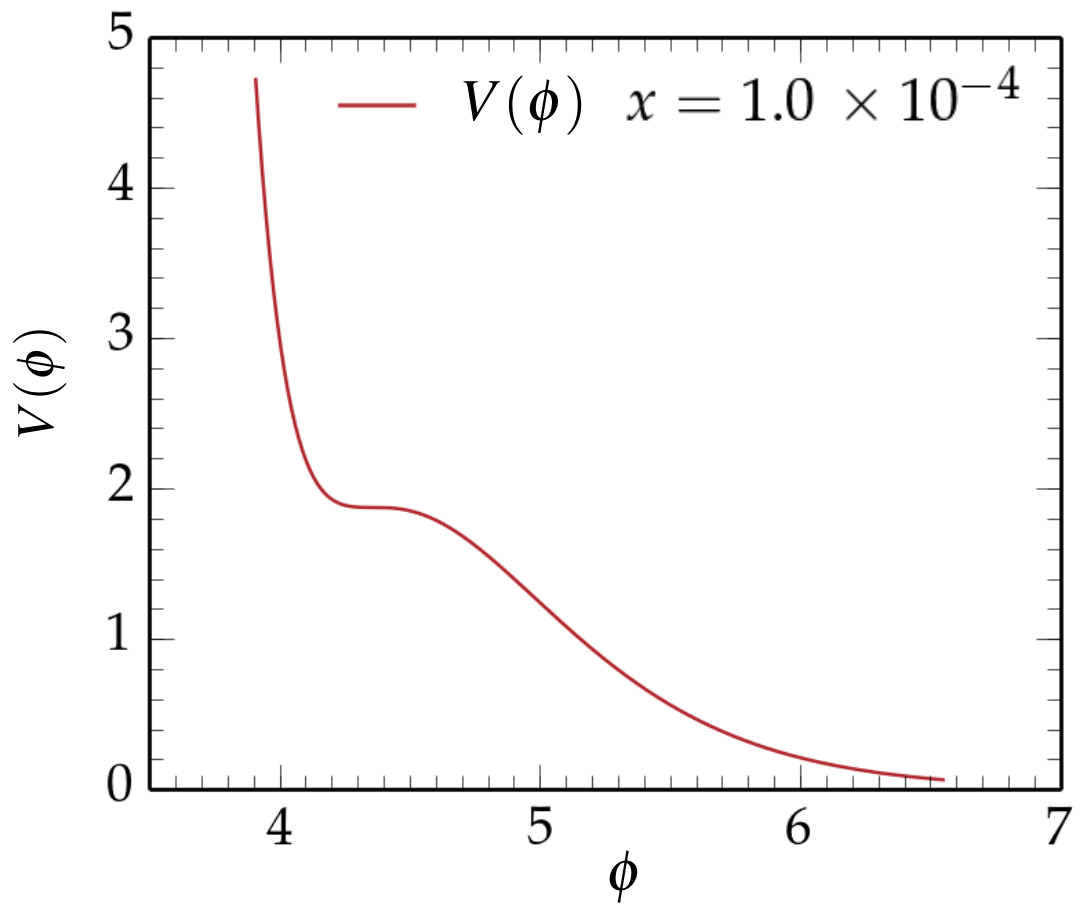}}
\caption{Scalar potential $V(\phi)$ for different values of $x$ giving an AdS, Minkowski or dS  vacuum.}
\label{figxparameter}
\end{figure}

In  section \ref{sectioninflation} we will study the inflationary possibilities from the above model. The inflaton will be identified with the canonically normalised $\phi/\kappa$, thus we express here the potential \eqref{Vlargelimit} in terms of the (dimensionless) inflaton $\phi$
\begin{equation}\label{potwitht}
V(\phi)\simeq -\frac{C}{\kappa^4}e^{-3 \sqrt{\frac 32} \phi} \left(\sqrt{\frac 32} \phi - 4 + q + \frac 32 \sigma e^{\sqrt{\frac 32} \phi}\right).
\end{equation}
We emphasize again that $\phi$ is dimensionless. In order to minimise and study the slow-roll parameters we compute the first two derivatives of $V$.  From \eqref{potwitht} we get
\begin{align}
V'(\phi)&= 3 \sqrt{\frac 32}\frac{C}{\kappa^4} e^{-3 \sqrt{\frac 32} \phi} \left(\sqrt{\frac 32} \phi + q -\frac{13}3  +\sigma e^{\sqrt{\frac 32} \phi}  \right),\\
V''(\phi)&=- \frac{27}2 \frac{C}{\kappa^4}e^{-3 \sqrt{\frac 32} \phi} \left(\sqrt{\frac 32} \phi +   q -\frac{14}3  +\frac 23 \sigma e^{\sqrt{\frac 32} \phi}  \right).
\end{align}
Solving $V'(\phi)=0$ leads to the two solutions
\begin{align}
&\phi_{-}=-\sqrt{\frac 23}\left(q-\frac{13}3+W_0\left(-e^{-x-1}\right)\right), \label{min}\\
&\phi_{+}=-\sqrt{\frac 23}\left(q-\frac{13}3+W_{-1}\left(-e^{-x-1}\right)\right), \label{max}
\end{align}
with $\phi_-$ the local minimum and $\phi_+$ the local maximum, with $\phi_-<\phi_+$.  $W_{0/-1}$ are the two branches of the Lambert function (or product logarithm) and $x$ is defined through the relation 
\begin{equation}
x\equiv q-\frac{16}3-\ln(-\sigma) \quad \leftrightarrow \quad \sigma=-e^{q-\frac{16}3-x}. \label{xdefinition}
\end{equation}
As mentioned above, it is clear from \eqref{min} and \eqref{max} that when $x$ is kept constant, varying $q$ shifts the local extrema.
The critical value $x_{c} \simeq 0.072132$ gives a Minkowski minimum, $i.e.$ with $V(\phi_-)=0$. The region $0<x<x_c$ gives a de Sitter (dS) minimum and $x>x_c$ gives an anti-de Sitter (AdS) one. The region $x<0$ corresponds to the case where the two branches of the Lambert function join and the potential loses its local extrema. The shape of the potential in the three regimes is shown in Figure \ref{figxparameter}.

\begin{figure}
\begin{minipage}{0.5\textwidth}
   \hspace{-1.5cm}\includegraphics[scale=0.35]{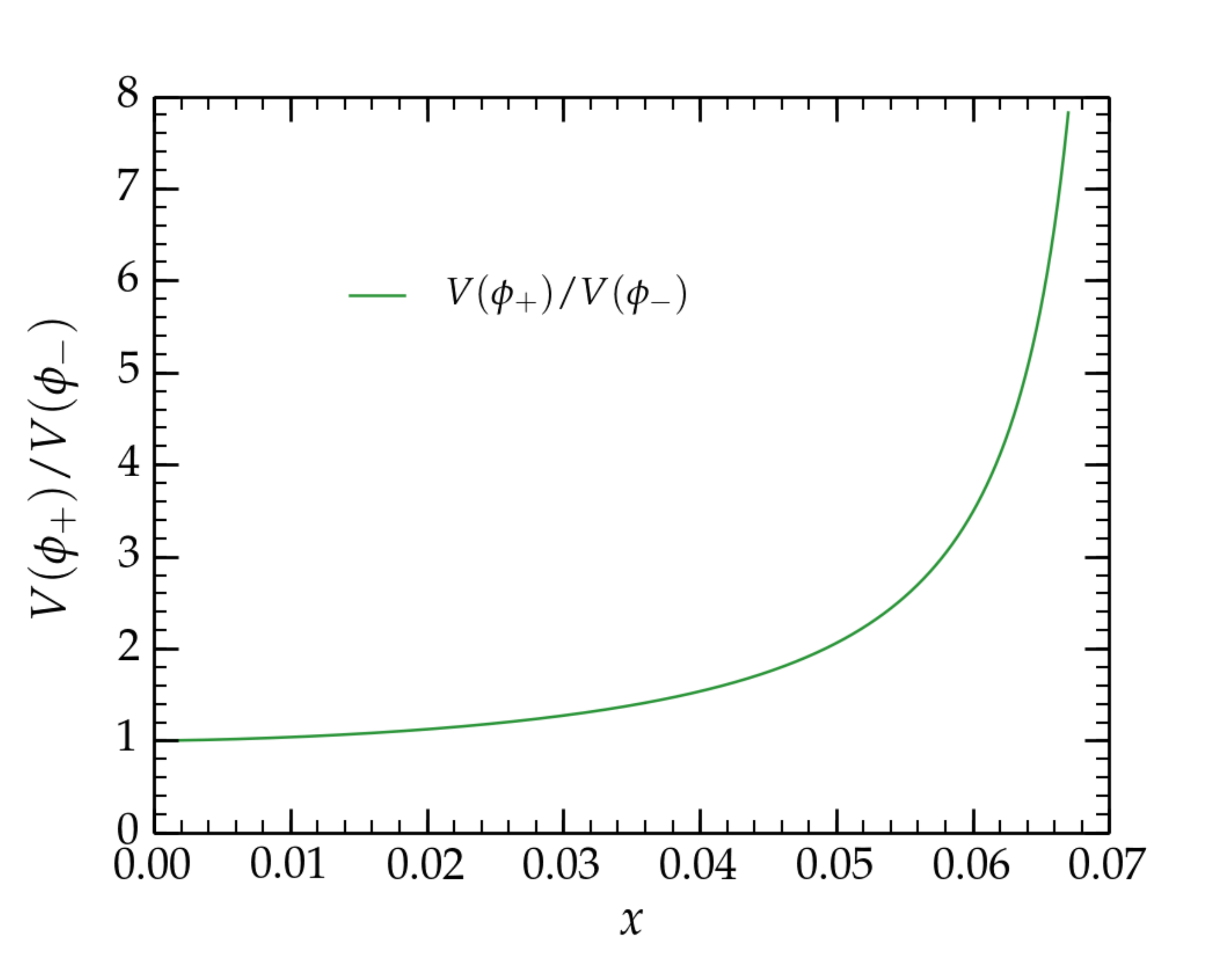}      
  \end{minipage}
    \hspace{0.6cm}
  \begin{minipage}{0.5\textwidth}
   \hspace{-1cm}\includegraphics[scale=0.35]{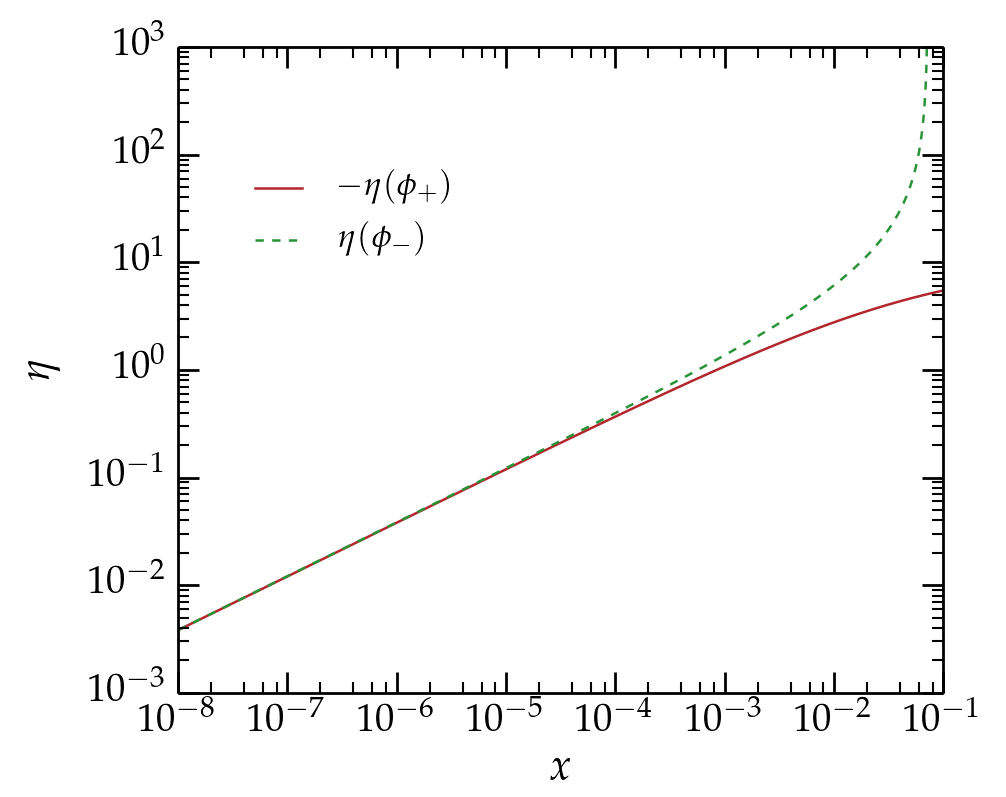}      
  \end{minipage}
  \caption{Ratio of the values of the scalar potential $V$ at the two extrema ({\it left panel}) and value of the slow-roll parameter $\eta_V$ at the two extrema ({\it right panel}), as functions of the parameter $x$.}
\label{Figratio}
 \end{figure}

 The values of the potential and its derivatives at the extrema can be derived from \eqref{min} and \eqref{max}:
 \begin{align}
&V(\phi_{-/+})=- \frac{C}{6\kappa^4} e^{-13+3q+ 3 W_{0/-1}(-e^{-x-1})}\left(2+3 W_{0/-1}(-e^{-x-1})\right), \label{Vminmax}\\
&V'(\phi_{-/+})= 0,\\
  &V''(\phi_{-/+})=\frac{9C}{2\kappa^4} e^{-13+3q+ 3 W_{0/-1}(-e^{-x-1})}\left(1+ W_{0/-1}(-e^{-x-1})\right). \label{V2minmax}
 \end{align}
 Firstly, from \eqref{Vminmax} we see that only the parameter $x$ determines the ratio between the values of the potential at the extrema. Indeed we get
 \begin{equation}\label{ratioV}
 \frac{V(\phi_{+})}{V(\phi_{-})}=\frac{\left(W_{0}(-e^{-x-1})\right)^3\left(2+3W_{-1}(-e^{-x-1})\right)}{\left(W_{-1}(-e^{-x-1})\right)^3\left(2+3W_{0}(-e^{-x-1})\right)}.
 \end{equation}
 This ratio is plotted in the left panel of Figure \ref{Figratio}.
 Secondly, we see that at the two extrema, the slow-roll parameter $\eta_V=\frac{V''}{V}$ also depends on $x$ only. It reads
  \begin{equation} \label{etaextrema}
\eta(\phi_{-/+})=\frac{V''(\phi_{-/+})}{V(\phi_{-/+})}=-9\frac{1+  W_{0/-1}(-e^{-x-1})}{\frac 23 + W_{0/-1}(-e^{-x-1})}.
 \end{equation}
Again, from \eqref{etaextrema} and \eqref{ratioV} we see that $x$ is the only important parameter of the model for the shape of the potential. From Figure \ref{Figratio} we see that as soon as $x\lesssim 0.05$ there is no scale separation anymore between the values of the potential at the two local extrema. 

\section{Inflation possibilities from the model} \label{sectioninflation}
\subsection{Generalities about inflation}

Inflation is characterized by an accelerated expansion of the Universe. In the standard Friedmann-Lema\^itre-Robertson-Walker (FLRW) metric parametrized by a scale factor $a$, an inflationary phase occurs when $\ddot{a}>0$. The dots denote derivatives with respect to the FLRW time $t$. We define the Hubble parameter
\begin{equation}
H(t)=\frac{\dot{a}}{a},
\end{equation}
and recall the Friedmann equations for an expanding Universe filled with a single scalar field
\begin{align}
&3H^2=\frac 12 \dot{\phi}^2+\kappa^2V(\phi) ,\label{H2}\\
&2\dot{H}=- \dot{\phi}^2, \label{dotH}
\end{align}
as well as the Klein-Gordon equation for the scalar field in FLRW background
\begin{equation}
\ddot{\phi}+3H\dot{\phi}+\kappa^2V'(\phi)=0~. \label{KG}
\end{equation}
We recall that our $\phi$ is dimensionless. Following the Hamilton-Jacobi method \cite{Salopek:1990jq}, we make a change of variable to take the inflaton $\phi$ as the time variable. This change of variable is given by rewriting $\dot{H}=\frac{dH}{d\phi}\dot{\phi}$ in equation \eqref{dotH}, leading to  
\begin{equation}
\frac{dH}{d\phi}=H'(\phi)=-\frac {1}2 \dot{\phi}~.
\end{equation}
Using \eqref{H2} and expressing $\dot{\phi}$ as a function of $H$ and $V$, we obtain
\begin{equation}
H'(\phi)=\mp \frac 1{\sqrt{2}}\sqrt{3H^2(\phi)-\kappa^2 V(\phi)}~. \label{dHdphi}
\end{equation}

\noindent 
We also use the exact slow-roll parameters defined as \cite{Liddle:2000cg}
\begin{equation}
\eta(\phi)=2 \, \frac{H''(\phi)}{H(\phi)}~, \qquad \epsilon(\phi)=-\frac{\dot{H}}{H^2}=2 \left(\frac{H'(\phi)}{H(\phi)}\right)^2~. \label{truesrparams}
\end{equation}
From the first expression of $\epsilon$ in \eqref{truesrparams}, we obtain
\begin{equation}
\frac{\ddot{a}}{aH^2}=1-\epsilon,
\end{equation}
so that $\epsilon<1$ is the natural criterion characterising inflation. The parameters defined in \eqref{truesrparams} are different from their slow-roll approximations expressed in terms of the potential only
\begin{equation}
\eta_V(\phi)= \frac{V''(\phi)}{V(\phi)}, \quad \epsilon_V(\phi)= \frac 12 \left(\frac{V'(\phi)}{V(\phi)}\right)^2, \label{srparams}
\end{equation}
with slow-roll limit $ \epsilon \underset{\epsilon\ll1}{\rightarrow }\epsilon_V, \,\, \eta\underset{\eta\ll1}{\rightarrow } \eta_V-\epsilon_V.$

When these two slow-roll parameters are small, $\dot{\phi}$ can be neglected in \eqref{H2} (and $\ddot{\phi}$ in \eqref{KG}) and we get the following Hubble constant slow-roll solution
\begin{equation}
H_{sr}(\phi)=\kappa \sqrt{\frac{V(\phi)}{3}}.
\end{equation}

In general (slow-roll or not), the number of e-folds $N$ before the end of inflation is defined by
\begin{align}
N&=\ln\left(\frac{a_{end}}{a}\right)=\int_t^{t_{end}} H dt= \int^{\phi_{end}}_{\phi} H \frac{d\phi}{\dot{\phi}} = -\frac{1}2 \int^{\phi_{end}}_{\phi} \frac{H}{H'}d\phi = \frac{1}{\sqrt{2}} \int_{\phi_{end}}^{\phi} \frac{d\phi}{\sqrt{\epsilon}} . \label{efoldsformula}
\end{align}
From equation \eqref{efoldsformula} we observe that in order to obtain the correct amount of e-folds near the minimum, $i.e.$ when $V'=0$, we would rather use definition \eqref{truesrparams} of $\epsilon$ rather than $\epsilon_V$ of \eqref{srparams}. Nevertheless, in our study, the definition \eqref{srparams} of $\eta_V$ is sufficient and is the one we will use (dropping the subscript $V$). Indeed, we only use this slow-roll parameter to determine when the modes exit the horizon. As we will see in the next sections, at this point $\epsilon_* \ll \eta_* \ll 1$ and thus, according to \eqref{srparams}, 
the slow-roll expression $\eta_V$ gives the correct estimate. 

We recall now that any model of inflation is constrained by the observed inhomogeneities in the energy density power spectrum, related to the temperature anisotropies of the cosmological microwave background (CMB) radiation, or to the number density of galaxy clusters. The three observational constraints are the following:
\begin{enumerate}
\item In order to solve the horizon problem, inflation must last for at least $60$ e-folds after horizon exit of the interesting modes.\footnote{This is true for high inflation scale. For lower energies the required numbered of e-folds decreases up to about 50 for TeV scale inflation.} Hence we must obtain 
\begin{equation}N_*=\int_{\phi_{end}}^{\phi_*} \frac{d\phi}{\sqrt{\epsilon}} \gtrsim 60.\label{N}
\end{equation} 
The star in $N_*$ or $\phi_*$ denotes the values of variables taken at horizon exit.
\item The observed spectral index $n_S$ measuring the deviation from a scale-invariant power spectrum, is related to the slow-roll parameters at horizon exit by
\begin{equation} 
n_S-1=2\eta_*-6\epsilon_*\simeq-0.04\,.
\label{ns}
\end{equation}
\item The spectral amplitude $\mathcal{A}_S$ induced by observations is
 \begin{equation}\mathcal{A}_S=\frac{\kappa^4V_*}{24\pi^2\epsilon_*}\simeq2.2 \times 10^{-9}.\label{As}\end{equation}
\end{enumerate}
 
\subsection{Hilltop inflation}\label{hilltopinflation}

The hilltop inflation scenario \cite{Linde:1981mu, Boubekeur:2005zm} emerged more than thirty years ago. The idea is the following: the inflaton starts rolling from a local maximum down to the minimum of the potential. In the vicinity of the local maximum, the slow-roll parameter $\epsilon$ is negligible while $\eta$ is determined by the observed spectral index. The horizon exit occurs near the maximum, and the $60$ remaining e-folds are obtained from there to a point before the minimum, where $\epsilon=1$ and inflation stops. As $\epsilon \rightarrow 0$ at the maximum, one can generate an infinite number of e-folds. In reality, the number of e-folds is dictated by the initial condition. The closest to the maximum the inflaton starts rolling down, the largest the number of e-folds is. 

The fact that the inflaton starts rolling from the maximum of the potential may be motivated if one considers that this maximum was related to a symmetry restoration point. At higher temperatures this point could have been a symmetric minimum of the potential, which became a maximum after spontaneous symmetry breaking occurring when temperature cooled down. Hence if the inflaton sits at a symmetric point at higher temperatures, it is natural to take initial conditions near the maximum.

In our model, according to \eqref{etaextrema} the values of the $\eta$ slow-roll parameter at the maximum only depends on $x$. Solving  \eqref{etaextrema} in order to have $\eta(\phi_{+})\simeq -0.02$ we find $x\simeq2.753 \times 10^{-7}$. This value can also be obtained graphically from the right panel of Figure \ref{Figratio}.

In order to study the possibility of hilltop inflation we then take  $x\simeq2.753 \times 10^{-7}$, $q=0$ and $C=1.0 \times 10^8$ and solve \eqref{dHdphi} numerically.\footnote{The $C$ coefficient being an overall scaling, its value is not important in the numerical studies. Hence we choose it, such that the potential is of order $1$. The true value of $C$ is determined by the amplitude \eqref{As} in the end.} This allows to find the Hubble parameter solution all along the inflation trajectory. Due to the shape of the potential (containing exponentials and linear terms), we had to use very high precision data types. This was achieved through a \verb|C++| code using \verb|cpp_dec_float| data types provided by the $\verb|multiprecision|$ package of the $\verb|boost|$ library. 

\begin{figure}[h]
   \includegraphics[scale=0.27]{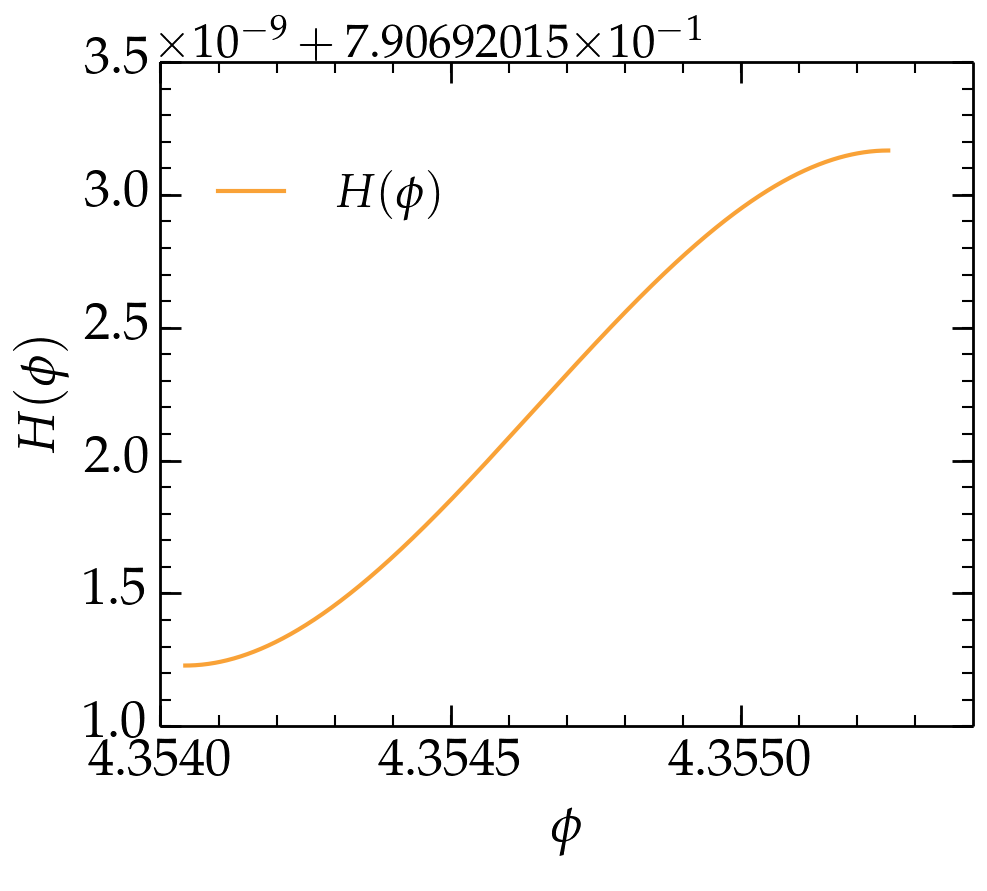}      
   \includegraphics[scale=0.27]{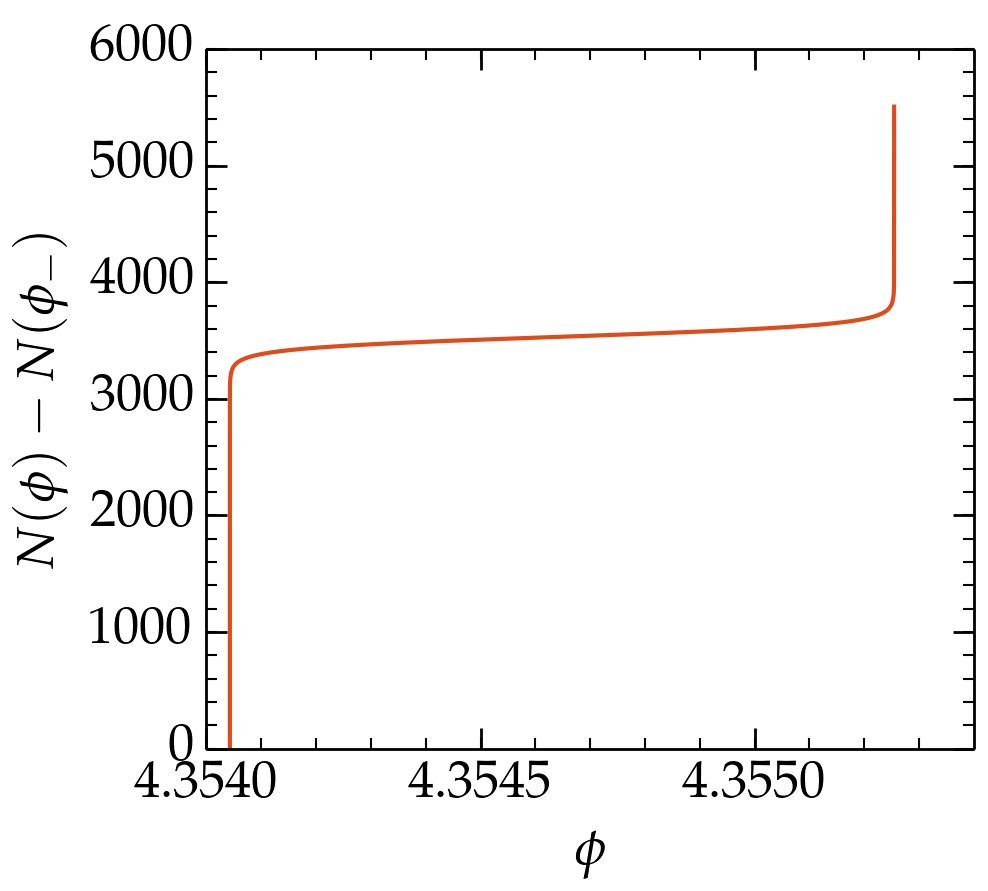}      
     \begin{minipage}{1\textwidth}
   \centering \includegraphics[scale=0.38]{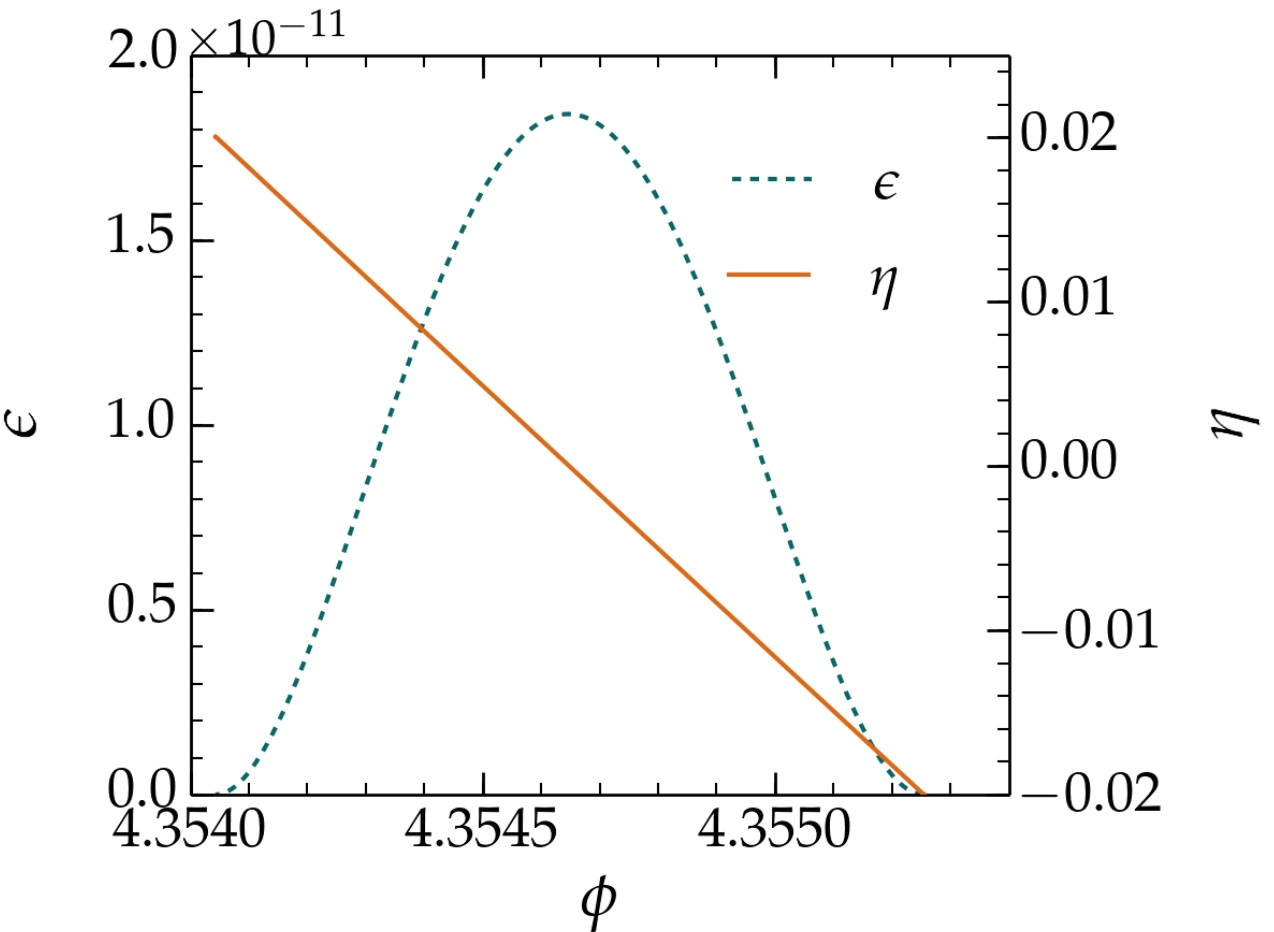}      
   \end{minipage}
\caption{Hubble parameter $H(\phi)$ ({\it top left}), number of e-folds $N(\phi)$ ({\it top right}) and slow-roll parameters $\epsilon$, $\eta$ ({\it bottom})  for $q=0$, $C=1\times 10^8$ and $x=2.753 \times 10^{-7}$.}
\label{plot1}
 \end{figure}

Plots of all the interesting parameters are shown in Figure \ref{plot1}. The horizontal axes show the values of the inflaton $\phi$. As the field starts from the maximum and goes towards the minimum, the arrow of time is from right to left ($i.e.$ for decreasing $\phi$.) 

We see from the bottom panel of Figure \ref{plot1} that the slow-roll parameters $\epsilon$ and $\eta$ stay small, hence the slow-roll regime holds all the way from the maximum down the minimum.  Since $\epsilon \ll 1$, inflation continues until the minimum and there is no natural criterion marking the end of inflation.  Of course, the value of the potential at the minimum being of order of the inflation scale, some new physics should be added to lower the potential near the actual cosmological constant. Nevertheless we see from Figure \ref{plot1} that there is a huge number of e-folds all along the inflationary trajectory, $i.e.$ between the would be horizon exit at $\eta_*=-0.02$ (near the maximum) and the minimum. Hence, this model cannot accommodate hilltop-inflation scenario because the constraints $N_*\simgt 60$ and $\eta_*=-0.02$ cannot be satisfied together by adjusting $x$, the only relevant parameter here.

  \subsection{Inflation around the minimum from the inflection point}\label{inflationminimum}
  
  \paragraph{General idea.} We now consider the case where the e-folds are obtained only near the minimum. This allows to alleviate the constraint $\eta(\phi_+)\simeq -0.02$.  We start with initial conditions near the maximum with no initial speed. We come back to this point at the end of the section. The inflationary phase corresponds to the inflaton rolling down its potential. As it goes from the maximum to the minimum, the second derivative $V''(\phi)$ changes sign and if $\eta(\phi_+)<-0.02$, it will pass through the value $\eta(\phi_*)=-0.02$ before the inflection point. The $x$ parameter of the model can then be chosen so that at least $60$ e-folds are obtained from this point to the end of inflation. 
   From the above argument we see that in order this scenario to correctly match the observational data, the initial position of the inflaton has to be higher than the inflection point, where $\eta$ is negative, so that $\eta=-0.02$ is taken at the horizon exit. 
      
\begin{figure}[h]
   \includegraphics[scale=0.27]{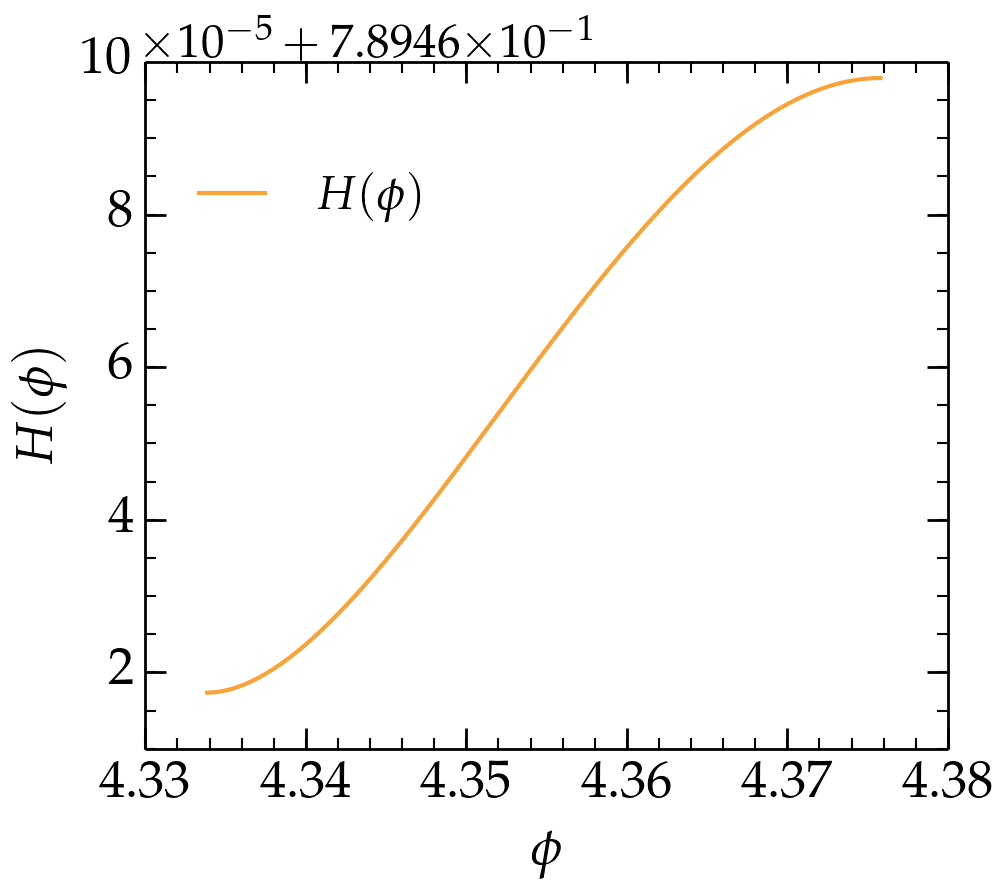}         \includegraphics[scale=0.27]{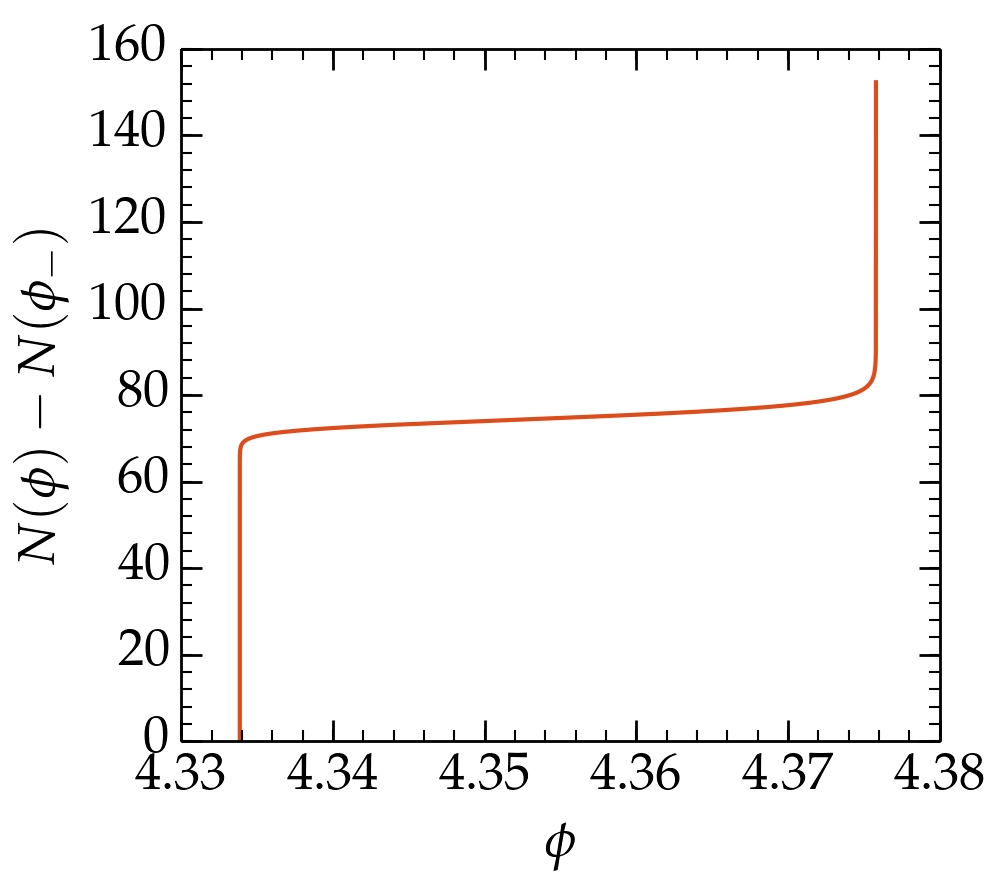}      
  \begin{minipage}{1\textwidth}
     \vspace{0.2cm}
\centering \includegraphics[scale=0.39]{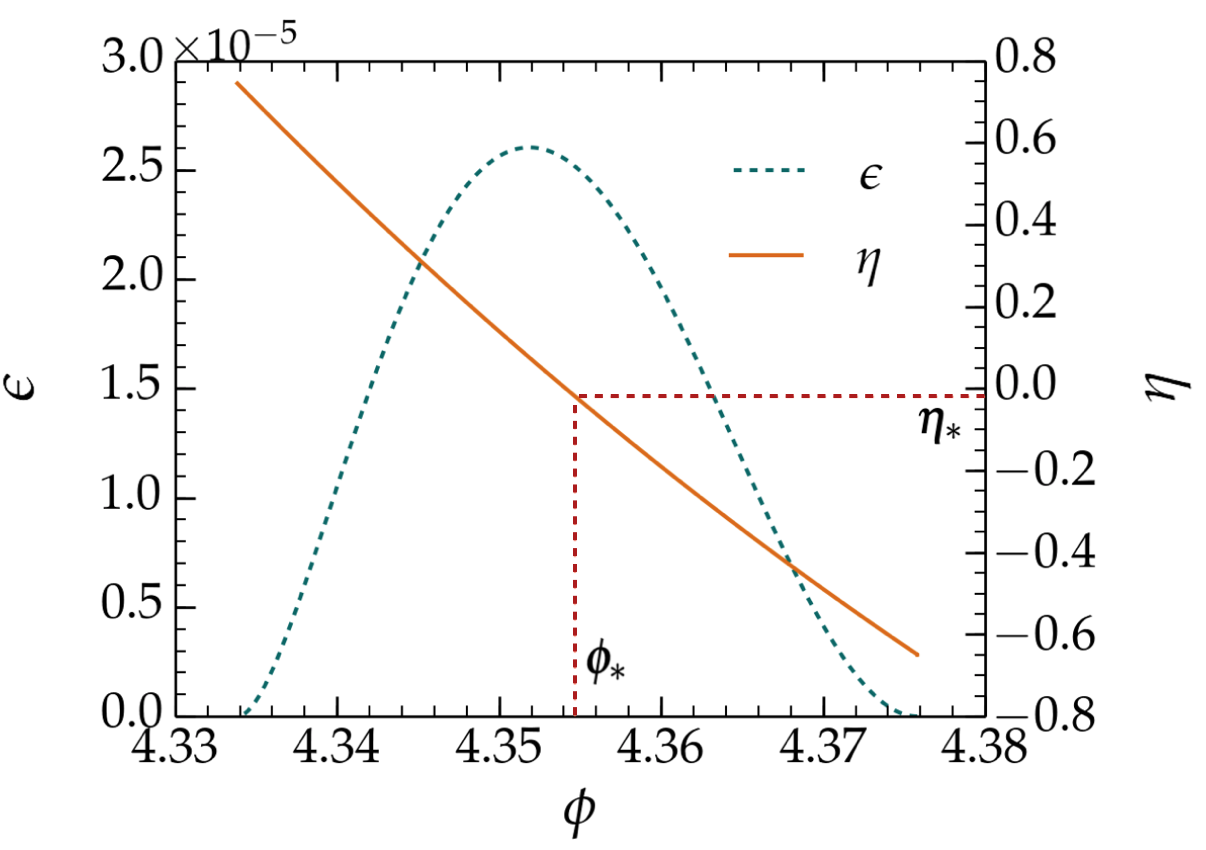}      
  \end{minipage}
\caption{Hubble parameter $H(\phi)$ ({\it top left}), number of e-folds $N(\phi)$ ({\it top right}) and slow-roll parameters $\epsilon$, $\eta$ ({\it bottom}), for $q=0$, $C=1\times 10^8$ and $x=3.3 \times 10^{-4}$. The dashed horizontal line shows the value $\eta=-0.02$.}
\label{plot2}
 \end{figure}

As in the hilltop case of section \ref{hilltopinflation}, we solved the evolution equation \eqref{dHdphi} numerically starting near the maximum with vanishing inflaton initial speed. We got $N_*\simeq60$ for  $x\simeq3.3\,\,10^{-4}$. In that case, $\phi_-=4.334$ and $\phi_+=4.376$. The e-folds are computed from the horizon exit $\phi_*\simeq 4.354$ at which $\eta(\phi_*)=-0.02$, to the minimum $\phi_-$. Is should be observed  that the corresponding inflaton field displacement 
\begin{equation} 
\Delta\phi\simeq 0.02~,
\label{deltaphi}
\end{equation}
is much less than one in Planck units, corresponding to small field inflation, compatible with the validity of the effective field theory.
We show the numerical solution in Figure \ref{plot2}. Again, the horizontal axes correspond to the inflaton $\phi$. As the field starts from the maximum and goes towards the minimum, the arrow of time is from right to left ($i.e.$ for decreasing $\phi$.)

In Figure \ref{plot2} we see that at the minimum neither $\epsilon$ nor $\eta$ is bigger than one ($\eta$ is however close to $1$). It is easy to understand from the plot of the $\epsilon$ parameter and formula \eqref{efoldsformula} that almost all e-folds are obtained near the minimum because $\epsilon$ is very tiny there. The vertical line shows the value of $\phi_*$ for which $\eta(\phi_*)=-0.02$. It is very close to the inflection point, hence the modes exit the horizon near (a bit before) the inflection point.

\paragraph{Study near the minimum.}   We wish to describe carefully what happens close to the minimum. Indeed, one has to check that the field goes on the other side of the minimum ($\phi<\phi_-$). We then expect that the field stops at a value $\phi_{stop}$ and goes back towards the minimum, starting its oscillation phase, usually related to the reheating period and the inflaton decay \cite{Abbott:1982hn,Albrecht:1982mp,Turner:1983he}. Nevertheless, one usually assumes tha the inflaton potential (almost) vanishes at the minimum. The inflaton decays into other particles and the cosmological constant stays then neglible in front of the radiation and matter densities for a sufficiently long period of time (assuming a solution to the cosmological constant problem).
  
   In our model, for the parameter $x\simeq 3.3\times 10^{-4}$ chosen here, there is no scale separation between the inflation scale $H_*$ (at horizon exit) and the scale at the minimum, see Figure \ref{Figratio}. Therefore the standard reheating scenario cannot  occur, because the potential energy of the inflaton (or equivalently the cosmological constant) remains important at the minimum. Hence, if nothing is modified in the model the energy density of the created particles stays small compared to the cosmological constant. We come back to this discussion in section \ref{newphysics}.

In the $\verb|C++|$ program used to solve numerically \eqref{dHdphi}, the field values are stored in \verb|cpp_dec_float| types of variable, available in the $\texttt{multiprecision}$ package of the \texttt{boost} library. These variables allow to store numbers of the form $a \times 10^{n}$ with a 100 digits precision on the coefficient $a$.
When we reach the minimum, the field evolves very slowly and if $\phi_{stop}-\phi_{-}$ is small, so is $H(\phi_{stop})-H(\phi_-)$  and the 100 digits precision is not enough to determine when the inflaton stops ($i.e.$ when $\partial H / \partial \phi=0$). In order  to bypass this difficulty, we expand the Hubble parameter $H$ around the slow-roll solution $H_{sr}$ by defining a new variable $\delta H$ through
\begin{equation}
H=H_{sr}+\delta H =\sqrt{\frac V3}+\delta H~.
\end{equation}
Replacing the first derivative of $\delta H$,
\begin{equation}
\delta H'=H'-H_{sr}'=H'-\frac{1}{2\sqrt{3}}\frac{V'}{\sqrt{V}}~,
\end{equation}
in \eqref{dHdphi}, one finds the new form of equation \eqref{dHdphi}
\begin{equation}
\delta H'=\pm \frac{1}{\sqrt{2}}\sqrt{6 H_{sr}\delta H}-\frac{1}{2\sqrt{3}}\frac{V'}{\sqrt{V}}~, \label{evolvdH}
\end{equation}
where we neglected the $\delta H^2$ term. The advantage of this new formulation is that now, even if $\delta H$ is small compared to $H_{sr}$, their values are not stored using the same coeffiicient and we do not have to store precisely $H=H_{sr}+\delta H$.
The numerical solution of the evolution equation \eqref{evolvdH} confirms that the inflaton indeed reaches $\phi_{stop}<\phi_-$, but stays very close. Due to the small values of the slow-roll parameters (see Figure \ref{plot2}) the inflaton is still in a slow-roll regime near the minimum and the oscillations are very slow. In fact we checked numerically that the number of e-folds during the first oscillations is greater than the one from $\phi_*$ to $\phi_-$. As mentioned earlier, this is easily understandable considering the fact that due to the large value of $V_-=V(\phi_-)$ (same scale as the inflation scale) the kinetic energy of the inflaton is not big enough to produce particles which would change significantly the equation of state of the Universe. We come back to this point in section \ref{newphysics}.
 
     \begin{figure}[h]
 \centering 
\includegraphics[scale=0.32]{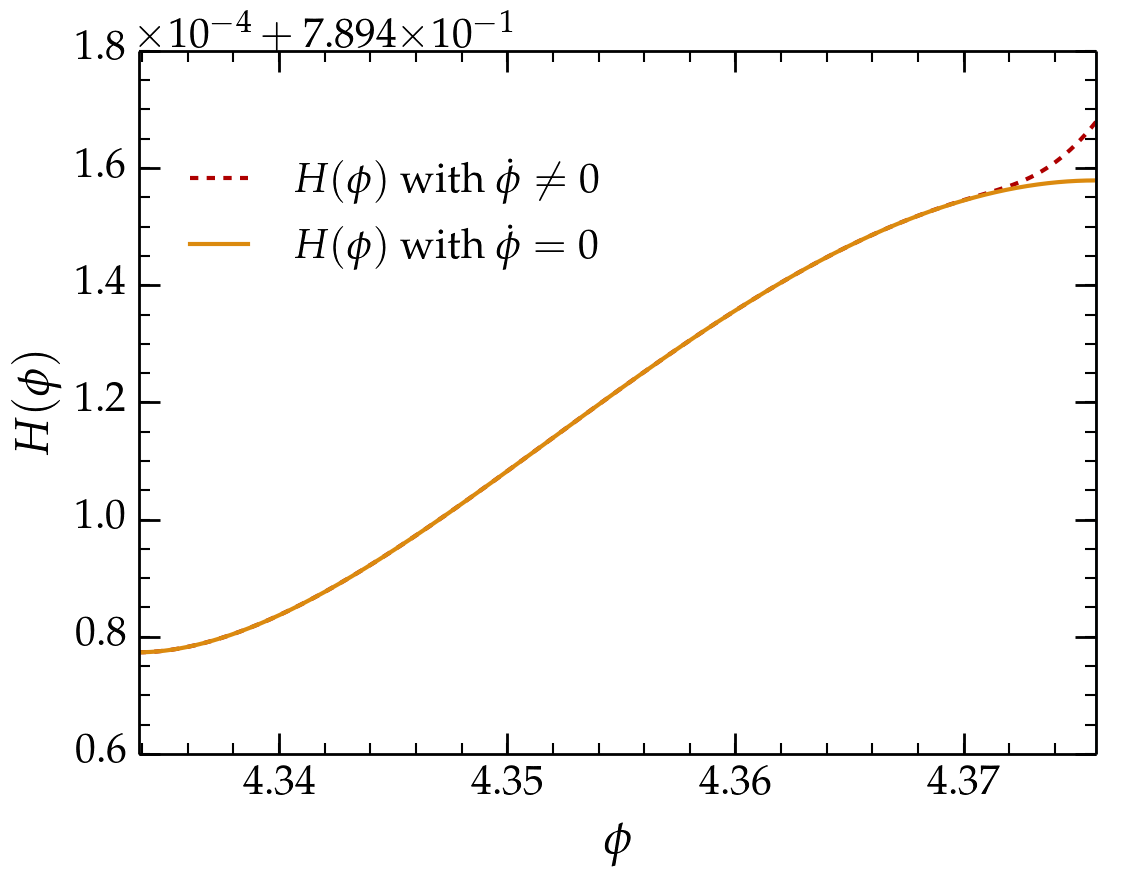} 
\caption{Hubble parameter solution for $x=3.3 \times 10^{-4}$, $q=0$, $C=1 \times 10^{8}$ with initial speed $\dot{\phi}=-2H'(\phi)=-2 \times 10^{-5} \kappa^{-1}$ ({\it red dashed line}), and with vanishing initial speed ({\it plain orange line}).}
\label{nonzerophidot}
 \end{figure}
 
 \paragraph{Initial conditions.} In the above study we started with the inflaton near the maximum with vanishing speed. In fact, one can change these initial conditions without altering  the conclusions of the study as long as the inflaton starts between the maximum and the inflection point with a relatively small speed. Indeed, the constraint that the inflaton starts higher that the inflection point comes from the fact that it has to cross the $\eta=-0.02$ point when rolling towards the minimum. Nevertheless, an argument of symmetry restoration similar to the one explained in section \ref{hilltopinflation} for the hilltop scenario motivates that the field starts near the maximum. In that case, if the initial speed stays relatively small, the inflaton is damped sufficiently near the maximum, such that the study does not change with respect with the case with vanishing initial speed. A solution for a non-zero initial speed is shown in Figure \ref{nonzerophidot}. If the initial speed is too large, equation \eqref{H2} shows that the major contribution to the Hubble parameter comes from the inflaton kinetic energy. As $V$ does not vary much from the maximum to the minimum, the inflaton only sees a flat potential until it reaches the wall at small $\phi$. In that case the previous study does not hold, and slow-roll inflation is not obtained.

 \subsection{Physical observables and theoretical parameters}\label{physicalparameters}
 
 In this section we study the implications of the  inflationary scenario described in section \ref{inflationminimum} to physical observables and we discuss the relation of the parameters of the model to those of the fundamental string theory.
 
 \paragraph{Inflation scale.} We see from Figure \ref{plot2} that when the modes exit the horizon,  the value of the slow-roll parameter related to the amplitude of primordial fluctuations is $\epsilon_*\simeq2.5 \times 10^{-5}$, implying a value for the ratio $r$ of tensor to scalar perturbations 
\begin{equation}
r=16 \epsilon_* \simeq 4\times 10^{-4}\,. 
\label{rpred}
\end{equation}
From \eqref{As} we deduce that
\begin{equation}
\kappa^4V_*=24\pi^2\epsilon_*A_S\simeq1.48 \times 10^{-11}. \label{Vstar}
\end{equation}
This constraint fixes the overall amplitude of the scalar potential. Indeed for the $x$ value of interest $V_*\simeq V(\phi_-)$ and we see from \eqref{Vminmax} that the value of the potential at the minimum reads
 \begin{align}
\kappa^4 V(\phi_{-})&=- \frac{C}{6} e^{-13+3q+ 3 W_{0}(-e^{-x-1})}\left(2+3 W_{0}(-e^{-x-1})\right)\equiv Ce^{3q}\times w(x), \label{vminf}
\end{align}
with  
\begin{equation}
w(x) =-\frac{1}{6}e^{-13+ 3 W_{0}(-e^{-x-1})}\left(2+3 W_{0}(-e^{-x-1})\right).
\end{equation}
 For the value of $x\simeq3.3 \times 10^{-4}$ realising the inflationary scenario described in section \ref{inflationminimum}, we obtain $w(x)\simeq 1.87 \times 10^{-8}$.
Together  with the constraint \eqref{Vstar}, equation \eqref{vminf} fixes the overall constant to
\begin{equation}
Ce^{3q}=\frac{\kappa^4V(\phi_-)}{w(x)}\simeq\frac{\kappa^4V_*}{w(x)}\simeq\frac{1.46 \times 10^{-11}}{1.87 \times 10^{-8}}\simeq7.81 \times 10^{-4}. \label{Ce3q}
\end{equation}
Note that for $q=0$ the value of $C$ is different from the one used in the plots of Figure \ref{plot2}, but as we explained in section \ref{dSminimumpot}, the overall constant just scales the potential and has no implication in the study of the inflation phase dynamics (in particular, it does not appear in the slow-roll parameters computation). 

From \eqref{Vstar} we deduce that the inflation scale is
\begin{equation}
H_*\simeq \kappa\sqrt{\frac{V_*}3} \simeq 2.2 \times 10^{-6} \kappa^{-1}\simeq 5.28 \times 10^{9}\,\,\, {\rm TeV}\,. 
\label{Hstar}
\end{equation}
 In the last equality we used the value of the Planck scale that we recall here:  $\kappa^{-1}\simeq2.4 \times10^{18} \,\,{\rm GeV}$.  As mentioned earlier, we observe from Figure \ref{plot2} that the value of the potential at the horizon exit and at the minimum are almost identical. Hence the positive value of the potential at the de Sitter minimum is given by $\kappa^4V_{dS}\simeq\kappa^4V_*$ and is way above the observed cosmological constant today. 

\paragraph{String parameters.} We now relate the parameters of the model to those of the underlying string theory and examine the constraints implied by the inflationary scenario described above. The string parameters are: $\xi$, $\gamma$, related to the quantum corrections, $d$ associated with the anomalous $U(1)$ charges of the 7-branes, and $\mathcal{W}_0$ the constant superpotential remaining after complex structure moduli and axion-dilaton stabilisation. 
For the sake of clarity, we write again the expressions for the first two in \eqref{xi} and \eqref{gammas}, resulting from string computations of quantum corrections~\cite{Antoniadis:2019rkh}:
\begin{equation}
\xi=-\frac{\zeta(3)}{4}\chi_{CY}, \qquad \gamma=-\frac 12 g_s T_0 \xi\,,
\end{equation} 
It follows that the parameters entering the scalar potential, introduced in \eqref{modelparameters}, read: 
\begin{equation}
q\equiv \frac \xi{2 \gamma}=-\frac{1}{g_s T_0}, \quad \sigma \equiv \frac{2d}{9 {\mathcal{W}_0}^2\gamma}<0, \quad C\equiv -3{\mathcal{W}_0}^2\gamma>0, \label{parametersbis}
\end{equation}
As already mentioned, note that in order to have $C>0$ we need $\gamma<0$ and hence a negative Euler number $\chi_{CY}$.

We have also defined the $x$ parameter by
\begin{equation}
x\equiv q-\frac{16}3-\ln(-\sigma) \,.
\end{equation}
From \eqref{min} we deduce that the volume at the minimum is a function of $q$ and $x$ only:
\begin{equation}
\mathcal{V}_-=\exp\left(\frac{3}{\sqrt{6}}\phi_-\right)=e^{-q} \times \exp\left(\frac{13}{3}+W_0\left(-e^{-x-1}\right)\right).
\end{equation}
Thus, for a given value of $x$, one obtains large volume for large (negative) $q$. In fact from \eqref{parametersbis}, $q$ is indeed negative for positive $T_0$, implying a surplus (locally) of D7-branes relative to orientifold O7-planes~\cite{Antoniadis:2019rkh}. Then large values of $q$ are reached as long as $g_s$ is small. Hence the weak coupling and large volume limits are related in a simple way. 

We now turn back to the string parameters $\mathcal{W}_0$ and $d$, which are partially fixed by the observational constraint through \eqref{Ce3q}. From the expressions \eqref{parametersbis},  the superpotential reads 
\begin{align}
\mathcal{W}_0^2&=-\frac{C}{3\gamma}
\simeq - \frac{2.64 \times 10^{-4}}{\gamma} e^{-3q}, \label{W0numerics}
\end{align}
whereas the $d$ parameter from the $U(1)$ D-terms of $D7$ branes is
\begin{align}
d&=\frac{9\mathcal{W}_0^2\gamma}{2} \sigma
=\frac{3C}{2} e^{q-\frac{16}{3}-x}\simeq \frac{3}2 \times 7.81 \times 10^{-4}  e^{-3q} \times e^{q-\frac{16}{3}-x} \simeq 5.65 \times 10^{-6}e^{-2q}\,.
\label{dnumerics}
\end{align}
We see from \eqref{W0numerics} that for values of $-\gamma$ around $10^{-2}$,  the value $\mathcal{W}_0\sim1$ is reached as soon as $-q\simgt 5$. On the other hand, from \eqref{dnumerics} we see that $d\sim1$ is reached for $-q\simgt 7$. 

We conclude that for  $-q={1}/(g_s T_0)$  not much greater than a few units, our inflationary model can be accommodated in the weak string coupling and large volume limits. This justifies that the large volume limit could safely be taken in the expressions \eqref{LVL} -- \eqref{VD} of the the scalar potential contributions $V_F$  and $V_D$.
Moreover,
the superpotential $\mathcal{W}_0$ and D-term coefficient $d$ take values of order one. In fact, $\mathcal{W}_0$ around unity can be naturally obtained from combinations of integer fluxes.

  \section{New physics around the minimum}
  
  In the section \ref{sectioninflation} we studied the inflation possibilities for the scalar potential of the type IIB string model described in section \ref{themodel}. We have seen in section \ref{inflationminimum} that it is possible to realise an inflationary period near the minimum of the potential, with horizon exit near the inflection point, by adjusting the parameter $x$ introduced in \eqref{xdefinition}.  We now address two important remaining questions concerning the stability of the minimum and the scenario for the end of the inflationary era.
  \subsection{Stability of the minimum}
  
  For the value of $x$ considered in section \ref{inflationminimum} in order to get an inflationary period, the values of the potential at the minimum and maximum are very close. Hence it is important to know if the inflaton can escape from the local minimum, or the false vacuum, and tunnel through the barrier of the potential before evolving classically towards the true minimum in the runaway direction at large field values. We recall that the shape of the potential for the value of the parameter $x$ giving an inflationary epoch is similar to the one shown in the right panel of Figure \ref{figxparameter}.

To evaluate the false vacuum stability we use the methods developed by Coleman et al. \cite{Coleman:1977py, Coleman:1980aw}. In order to keep their conventions, we will use the dimension-full inflaton 
\begin{equation}
\varphi=\frac{\phi}{\kappa}.
\end{equation}
This definition of the inflaton field $\varphi$ will be considered only in the current section.
 We are interested in the probability that the inflaton tunnels from the false vacuum at $\varphi_-=\kappa^{-1}\phi_-$ to the true vacuum located at the runaway direction $\varphi=+\infty$. Following the Coleman-de Luccia  (CdL) argument \cite{Coleman:1980aw}, this probability is expressed as a decay rate per unit volume and time by 
\begin{equation}
\Gamma=Ae^{-B}, \qquad {\rm with} \qquad  B=S_E(\varphi)-S_E(\varphi_-), \label{tunnelrate}
\end{equation}
where $S_E$ is the Euclidean tunnelling action to minimise. Here $\varphi$ denotes the instanton solution of the scalar field action. (For a recent review on vacuum stability, see~\cite{Markkanen:2018pdo}). For a scalar field coupled to gravity the Euclidean action reads
\begin{equation}
S_E=\int d^4x \sqrt{g}\left(-\frac 1{2\kappa^2} R + \frac 12 \partial_{\mu}\varphi\partial^{\mu}\varphi + V(\varphi)\right). \label{actionE}
\end{equation}

Following the strategy of CdL, one looks for a solution with an $O(4)$ symmetry. Such a Euclidean space-time can be described by the following metric
\begin{equation}
ds^2=d\rho^2+\chi(\rho)^2(d\Omega_3)^2, \label{metricE}
\end{equation}
were $(d\Omega_3)^2$ is the metric of the unit 3-sphere. The Euclidean scale factor $\chi(\rho)$ gives the curvature of the 3-sphere at given $\rho$. The Euclidean field equations for the scalar field $\varphi$ and the scale factor $\chi$ are 
\begin{align}
&\varphi''+3\frac{\chi'}{\chi}\varphi'=\frac{\partial V}{\partial \varphi}, \label{phiE}\\
&\chi''=-\frac{\kappa^2}{3}\chi\left(\varphi'^2+V(\varphi)\right) \qquad \left({\rm or } \, \,\,\, \chi'^2=1+\frac{\kappa^2}{3} \chi^2\left(\frac{1}2 \varphi'^2-V(\varphi)\right) \,\,\right)\label{scaleE} 
\end{align}
The prime denotes, in this section, derivative with respect to $\rho$ which plays the role of the time variable. 
Using the $O(4)$ symmetry of \eqref{metricE} and the field equations \eqref{scaleE}, one can rewrite the Euclidean action \eqref{actionE} as
\begin{equation}
S_E=2\pi^2\int d\rho \left( \chi^3\left(\frac12\varphi'^2+V\right)+\frac 3{\kappa^2} \left(\chi^2\chi''+\chi\chi'^2-\chi\right)\right)=-2\pi^2\int d\rho \chi^3(\rho) V\left(\varphi(\rho)\right). \label{actionE2}
\end{equation}

The two simplest solutions to \eqref{phiE} -- \eqref{scaleE} are the ones where the field $\varphi$ sits at an extremum of the potential $V$. These solutions read
\begin{equation}
\varphi(\rho)=\varphi_{\pm}, \quad \chi(\rho)=\frac{1}{H_{\pm}} \sin(H_{\pm} \rho), \quad {\rm with} \quad H_{\pm}=\kappa\sqrt{\frac{V(\varphi_{\pm})}{3}}, \label{dSsol}
\end{equation}
and are defined for $\rho \in [0, H_{\pm}^{-1}\, \pi]$. Once plugged back in \eqref{metricE}, the solution for $\chi(\rho)$ in \eqref{dSsol} gives simply the four sphere metric, which is the Euclidean extension of de Sitter space-time with $H=H_{\pm}$, obtained by analytic continuation of the real time to the Euclidean time $\rho$ \cite{Hawking:1973uf}. The solution with the inflaton sitting at the top of the barrier, $\varphi(\rho)=\varphi_+$ is related to the Hawking-Moss (HM) instanton \cite{Hawking:1981fz}. We come back to this solution later. We compute the action of solution \eqref{dSsol} through equation \eqref{actionE2}. It reads
\begin{equation}
S_E(\varphi_{\pm})=-2\pi^2\int d\rho\frac{1}{H_{\pm}^3}\sin^3(H_{\pm}\rho) V(\varphi_{\pm})=-\frac{24\pi^2}{\kappa^4V(\varphi_{\pm})}. \label{actionsdS}
\end{equation}

\begin{figure}[h] 
\makebox[\textwidth]{\hspace{-.3cm} \includegraphics[scale=0.32]{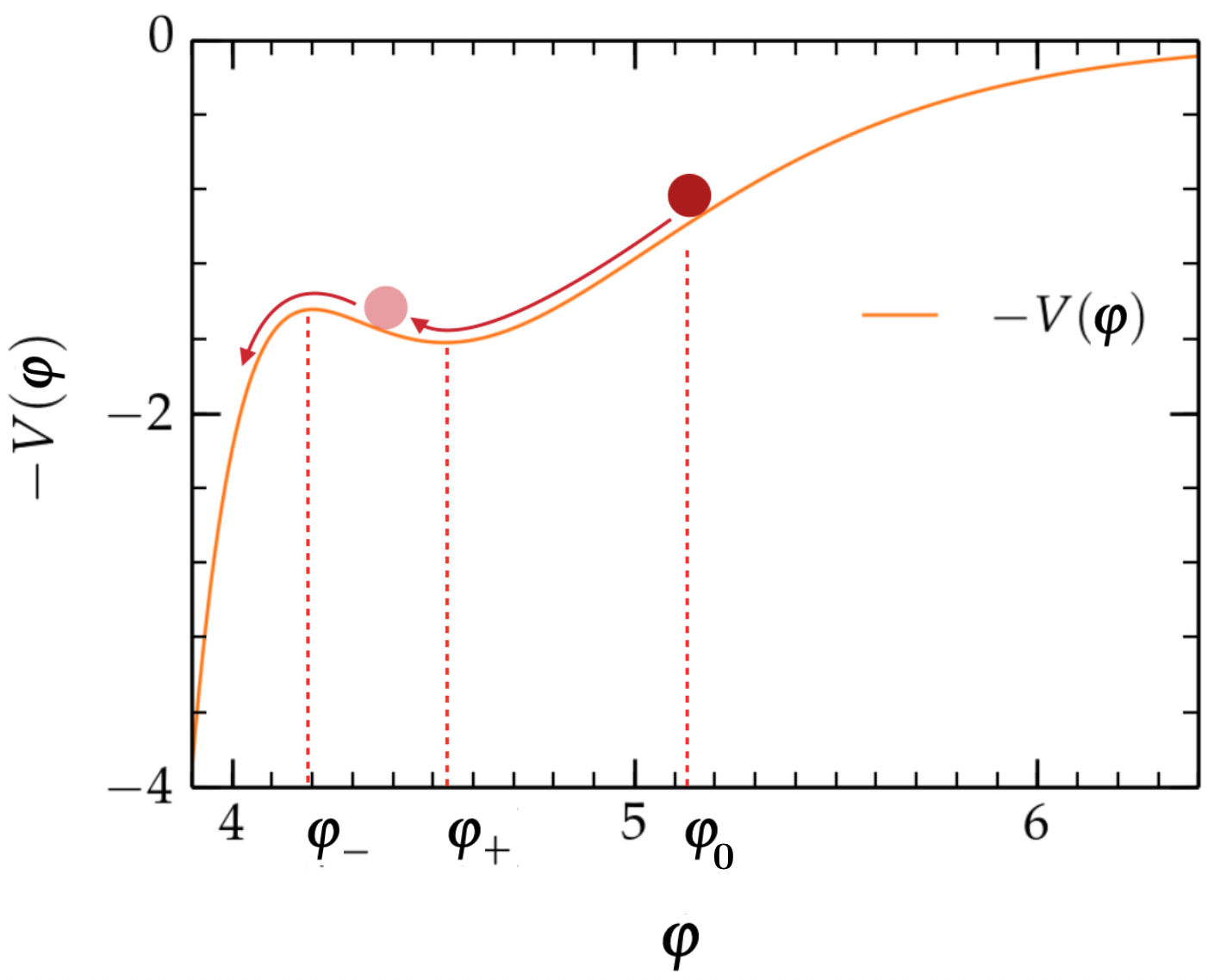}\hspace{0.2cm}
\includegraphics[scale=0.32]{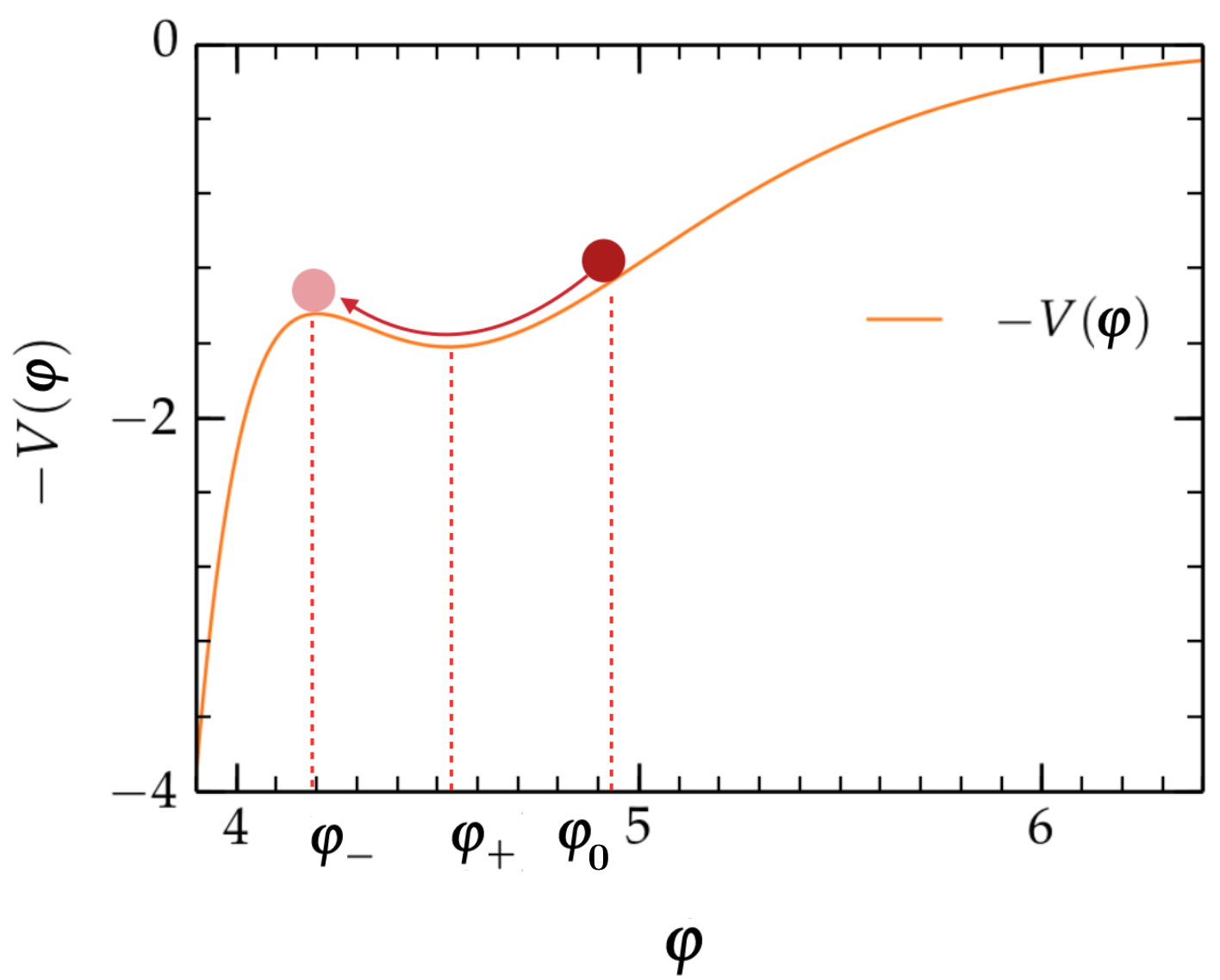}}
\caption{Point-like particle rolling down the inverted potential $-V(\varphi)$ from an initial position $\varphi_0$. An overshooting case is shown ({\it left panel}) as well as the CdL instanton solution ({\it right panel}) for which the field stops exactly at the maximum $-V(\varphi_-)$. }
\label{figintertedV}
\end{figure}

More complex solutions to \eqref{phiE} -- \eqref{scaleE}, referred as CdL instantons, can be found by imposing the following boundary conditions
\begin{equation}
\varphi'(0)=0, \quad \chi(0)=0, \quad \chi'(0)=1, \quad \varphi(\rho_f)=\varphi_-\,, 
\label{boundaryconditions}
\end{equation}
where the final time $\rho_f$ is determined through $\chi(\rho=\rho_f)=0$. At that time one should get $\varphi'(\rho_f)=0$. As one can see, the value $\varphi_0=\varphi(0)$ is not specified for the CdL instanton. 
Similarly to what was described originally in the case without gravity \cite{Coleman:1977py}, the boundary conditions \eqref{boundaryconditions} associated with the field equations \eqref{phiE} -- \eqref{scaleE} correspond to the classical problem of a field evolving in the reverse potential $-V$ and subject to a friction force proportional to $3\chi'/\chi$. Starting from an $a \, priori$ unknown position $\varphi_0$, it rolls down to the local minimum $-V(\varphi_+)$, passes it, climbs the hill and stops exactly at the local maximum $-V(\varphi_-)$. One has to find the value of $\varphi_0$ for which the field stops exactly at $\varphi_-$ with no speed. Figure \ref{figintertedV} sketches the situation of a point-like particle rolling down the inverted potential $-V(\varphi)$. Going back to the real time picture, $\varphi_0$ can be interpreted as the position reached by the scalar field after it tunneled the barrier from the valse vacuum $\varphi_-$. Then, from this point, the field evolves classically until the true (runaway) minimum at infinity.

Exact analytical solutions for the CdL instantons are difficult to find in general. They have been studied in the case without gravity for simple triangular or squared potentials \cite{Lee:1985uv}\cite{Jensen:1983ac}. Analytical solutions were also presented in the original papers \cite{Coleman:1977py}\cite{Coleman:1980aw} in the famous ``thin-wall'' approximation. This limit nevertheless demands that the false and true vacua, $V(\varphi_-)$ and $V(\infty)=0$, are very close with respect to the height of the barrier $\Delta V=V(\varphi_+)-V(\varphi_-)$. However, this approximation does not hold for the case we are studying, as can be seen for instance from the right panel of Figure \ref{figxparameter}.

CdL instantons can also be searched numerically following the undershooting, overshooting method proposed initially in the original paper without gravity \cite{Coleman:1977py}. The idea is very simple: we start with any value $\varphi_0$ and solve numerically \eqref{phiE} -- \eqref{scaleE} with initial conditions \eqref{boundaryconditions}. If the solution overshoots, $i.e.$ if the field continues rolling after having reached the local maximum  $-V(\varphi_-)$ of the reversed potential,  we start again with a new initial position $\varphi_0$ closer to the minimum $-V(\varphi_+)$. If the solution undershoots, $i.e.$ if the field does not reach the local maximum  $-V(\varphi_-)$, we start with a new $\varphi_0$ a bit further from the minimum $-V(\varphi_+)$. We repeat this operation until we find the initial value $\varphi_0$ for which the field stops exactly at the maximum of the reversed potential, which is the false vacuum.  We sketch the overshooting situation on the left panel of Figure \ref{figintertedV} and the CdL instanton solution on the right panel.

\begin{figure}[h] 
\makebox[\textwidth]{\includegraphics[scale=0.3]{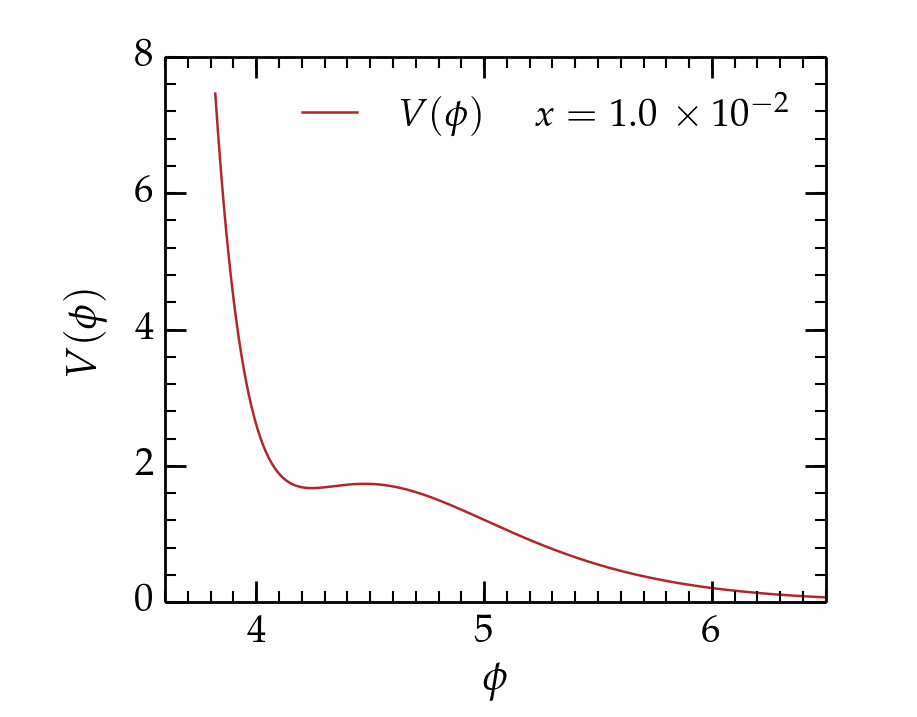}\hspace{-0.7cm}
 \includegraphics[scale=0.3]{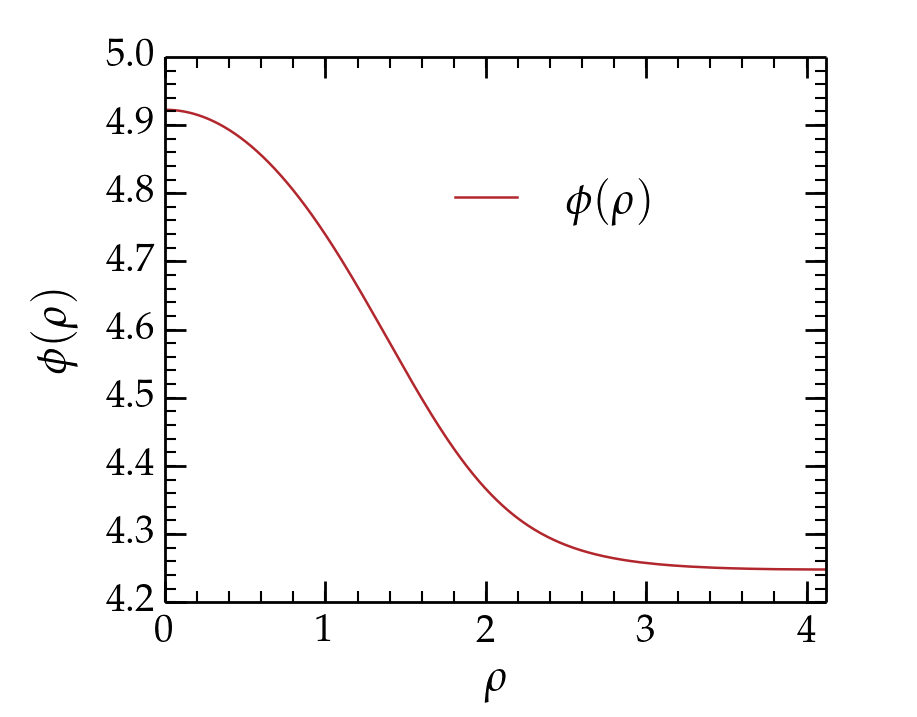}\hspace{-0.5cm}
\includegraphics[scale=0.3]{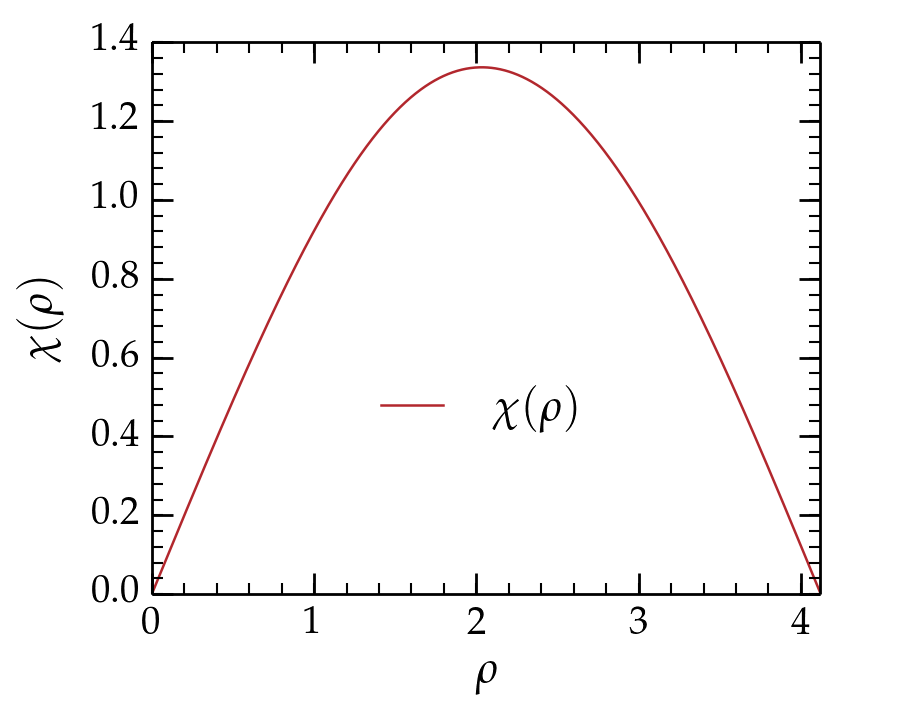}}
\caption{Scalar potential $V(\phi)$ for $x=1.0 \times 10^{-2}$ ({\it left}) and the CdL bounce solution functions $\phi(\rho)$ ({\it middle}) and $\chi(\rho)$ ({\it right}). The initial field value is $\varphi_0=4.9224\kappa^{-1}$ and the final Euclidean time is $\rho_f=4.1222$.}
\label{figbounce}
\end{figure}

In Figure \ref{figbounce}, we show the CdL instanton solution obtained with this method for the value of the parameter $x=1.0 \times 10^{-2}$. In this case, the field values at the two extrema are $\varphi_-= \phi_-\kappa^{-1}=4.2446 \kappa^{-1}$ and $\varphi_+= \phi_+\kappa^{-1}=4.4756\kappa^{-1}$, while the initial value is $\varphi_0= \kappa^{-1}\phi_0=4.9224\kappa^{-1}$.

\begin{figure}[h]
 \centering 
\includegraphics[scale=0.42]{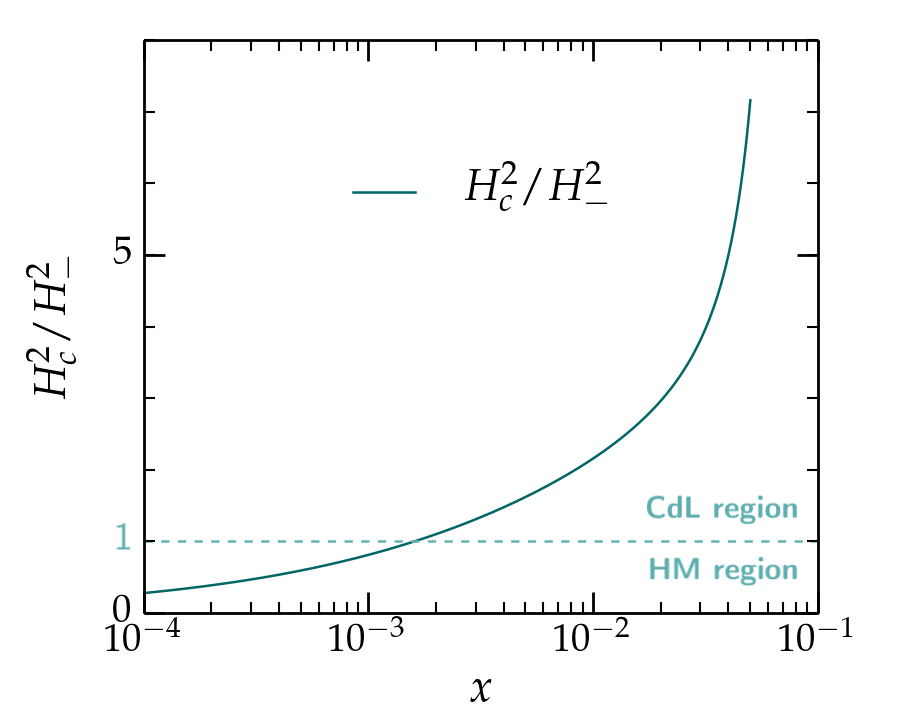} 
\caption{Ratio of the critical Hubble parameter $H_c^2$ and the Hubble parameter at the minimum $H_-^2=\kappa^2V(\varphi_-)/3$. The value $H_-=H_c$ ({ horizontal dashed line}) determines the existence and domination of the CdL instanton over the HM instanton.}
\label{figCdLorHM}
 \end{figure}

We tried to obtain a solution similar to the one of Figure \ref{figbounce} for $x=3.3 \times 10^{-4}$, which satisfies the observational constraints of the inflationary epoch, as explained in section \ref{inflationminimum} . Nevertheless we were not able to find a numerical solution satisfying the boundary conditions. In fact, if the potential barrier is too flat, the existence of CdL instanton is not guaranteed anymore. A criterion for the existence of CdL instanton has been established in the past following various arguments \cite {Jensen:1983ac} \cite{Balek:2003uu}. It demands that the scale factor at the minimum $H_-$ stays below a critical value $H_c$ defined by
\begin{equation}
H_c^2=-\frac{V_{\varphi\varphi}(\varphi_+)}{4}-\frac{\kappa^{2}\Delta V}{3}, \label{H_c}
\end{equation}
where $V_{\varphi\varphi}= {\partial^2 V}/{\partial \varphi ^2}$ and as above, $\Delta V=V(\varphi_+)-V(\varphi_-)$ is the height of the barrier. When $H_-$ approches $H_c$, the potential is flat near the maximum hence the $\Delta V$ contribution can be neglected in front of that of $V_{\varphi \varphi}(\varphi_+)$ in \eqref{H_c}. In this case, we can  use relations \eqref{Vminmax} and \eqref{V2minmax} to express the ratio 
\begin{equation}
\frac{H_c^2}{H_-^2}\simeq-\frac{3}{4} \frac{V_{\varphi\varphi}(\varphi_+)}{\kappa^2V(\varphi_-)}=\frac{81 W_0(-e^{-x-1})^3\left(1+W_{-1}(-e^{-x-1})\right)}{4W_{-1}(-e^{-x-1})^3\left(2+3W_0(-e^{-x-1})\right)}. \label{ratioHcritical}
\end{equation}
For $H_->H_c$ the existence of the CdL instantons is not guaranteed anymore. Figure \ref{figCdLorHM} shows the ratio of the Hubble scales in \eqref{ratioHcritical}.  In fact, the Hubble scale critical value $H_c$ also marks the dominance of the Hawking-Moss instanton \cite{Hawking:1981fz} solution. As explained before, this solution is the one for which the inflaton stays at the top of the barrier $\varphi(\rho)=\varphi_+$. This solution can be seen as describing the inflaton going above the potential barrier instead of properly tunnelling. As mentioned after \eqref{dSsol} such solutions always exist, but when $H_-<H_c$ their action is higher than the CdL solutions and give thus negligible contribution to the tunnelling rate \eqref{tunnelrate}.

In Figure \ref{figCdLorHM}, we see that  $H_->H_c$ holds for $x=3.3 \times 10^{-4}$, which explains that we are not able to find the standard CdL instanton. Hence in this case only the HM instanton contribute to the tunnelling rate. The tunnelling coefficient $B$ introduced in \eqref{tunnelrate} is then computed from \eqref{actionsdS} and reads
\begin{equation}
B=S_E(\varphi_+)-S_E(\varphi_-)=-\frac{24\pi^2}{\kappa^4V(\varphi_+)}+\frac{24\pi^2}{\kappa^4V(\varphi_-)} \simeq \frac{8\pi^2\Delta V}{3H_*^4}.
\end{equation}
In the last equality, $\Delta V$ is the height of the barrier and $H_*$ is the inflation scale, $i.e.$ the Hubble parameter when the modes exit the horizon. We recall that since in our model the potential is almost flat along the inflationary trajectory, we have $H_-\simeq H_+ \simeq H_*$. For $x=3.3 \times 10^{-4}$ we find $\Delta V\simeq 2.0 \times 10^{-4}V_*$ and from \eqref{Vstar} -- \eqref{Hstar} we deduce
\begin{equation}
B\simeq3.3 \times 10^9, \qquad i.e.\qquad  \Gamma=Ae^{-3.3 \times 10^9}.
\end{equation}
It follows that the decay rate of the local minimum is extremely small and the vacuum is practically stable.

\subsection{End of inflation: new physics around the minimum}\label{newphysics}

As established in the section \ref{inflationminimum}, having at least 60 e-folds near the minimum constrains the energy of the dS vacuum. This value is very much greater than the observed value today, hence this dS vacuum cannot be the true vacuum of the theory. Indeed, with such a big value, the Universe would continue expanding and never reach the standard cosmology with radiation and matter domination eras. 
Hence we see that in our model the cosmological constant problem arises naturally coupled to inflation. Of course we will not tackle this problem here, but show that taking it into account through the introduction of new physics near the minimum of our potential brings in a natural scenario for the end of the inflation epoch. This relates our model to the hybrid inflation proposal \cite{Linde:1993cn}, where a second field is added to the model. This so-called ``waterfall'' field adds another direction to the scalar potential. If falling into this direction becomes favorable at a certain point of the inflaton trajectory, this immediately ends the inflation era and the theory reaches another minimum at a different energy scale.

Below we describe a toy model of hybrid inflation adapted to our model. Consider the following Lagrangian for the inflaton $\phi$ and the extra waterfall field $S$
\begin{equation}
\mathcal{L}(\phi,S)=(\partial_{\mu}\phi)^2+(\partial_{\mu}S)^2-V(\phi)-V_S(\phi,S). \label{hybridpot}
\end{equation}
$S$ can be seen as a Higgs-like field undergoing symmetry breaking at an intermediate energy scale located between the inflation scale and the minimum of the potential. In  equation \eqref{hybridpot}, $V(\phi)$ is the inflaton potential studied previously and $V_S(\phi,S)$ an additional part containing the dependence in $S$ and its coupling with $\phi$. We express this second contribution in the following way
\begin{equation}
V_S(\phi,S)=\frac 12\left(-M^2+f(\phi)\right)S^2+\frac \lambda 4S^4,\label{Vsfs} 
\end{equation}
with $f(S)$ some function of $S$.
Depending on the sign of  its effective squared mass ${m_S}^2=-M^2+f(\phi)$, the  waterfall field $S$ stays in two separate phases. When ${m_S}^2>0$, $i.e.$ $M^2< f(\phi)$, the minimum in the $S$-field direction is at the origin 
\begin{equation}
\langle S \rangle=0, \quad {\rm when} \quad {m_S}^2=-M^2+f(\phi)>0\,,
\end{equation}
and the extra contribution to the scalar potential vanishes
\begin{equation}
V_S(0)=0\,.
\end{equation}
When the mass of $S$ becomes tachyonic, a phase transition occurs and the new vacuum is obtained at a non-vanishing $S$ vacuum expectation value 
\begin{equation}
\label{newmin}
\langle S \rangle=\pm \frac{\lvert{m_S}\rvert}{\sqrt{\lambda}}\equiv\pm{v}, \quad {\rm when} \quad {m_S}^2=-M^2+f(\phi)<0.
\end{equation}
The value of the potential $V_S$ at the minimum of this broken phase is
\begin{equation}
V_S(v)= -\frac{{m_S}^4}{4\lambda}<0.
\end{equation}
 
We can choose $f(\phi)$ such that during the phase of inflation, when the field $\phi$ rolls down the potential, we stay in the symmetric phase and the $S$ field is stabilised with a vanishing vacuum expectation value and a large positive mass. The inflation phase is then equivalent to the one field model studied in section \ref{inflationminimum}.

   \begin{figure}[h!]
 \centering 
\includegraphics[scale=0.6]{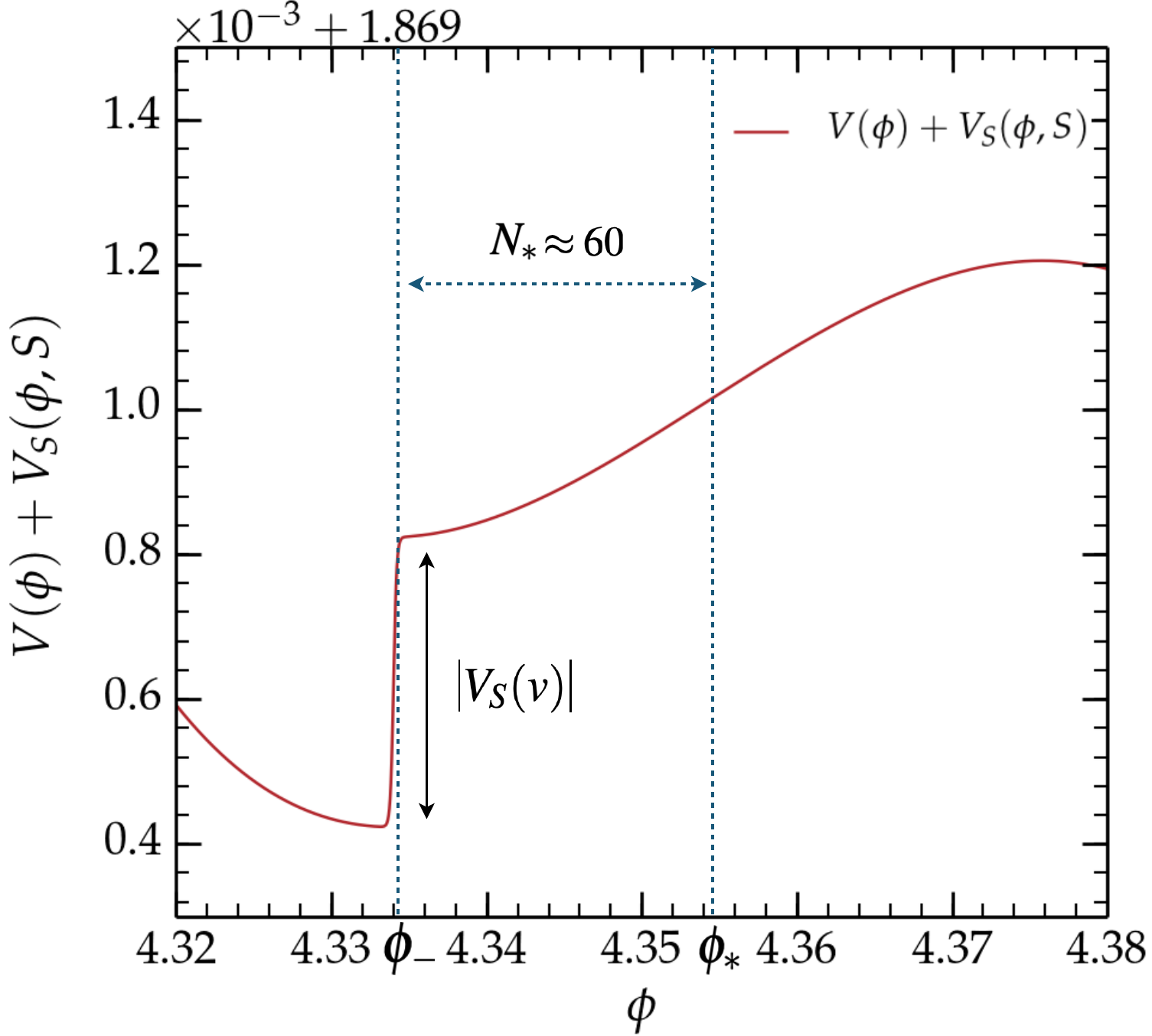} 
\caption{Schematic plot of an hybrid inflation scalar potential $V(\phi)+V_S(\phi,S)$ including a waterfall contribution from an extra field $S$}
\label{figwaterfall}
 \end{figure}

If near the minimum ${m_S}^2<0$, a phase transition occurs and the $S$ field goes to its value \eqref{newmin} at the new minimum. This amounts to change the potential $V(\phi)$ near the minimum, by a constant  $V_{down}=V_S(v)<0$. The effect of such a downlift is double: it decreases the value of the cosmological constant and if the waterfall direction is steep enough, it gives a natural criterion to stop inflation ($\epsilon>1$). We show in Figure \ref{figwaterfall} a schematic plot of such a potential.

In the string theory framework described in this paper, a waterfall field could be realized by an open string state at the intersection of the D7-branes stacks. 
Its mass would receive a supersymmetry breaking contribution from the configuration of the intersection angles and a supersymmetric one, 
from a separation in the transverse directions. The former can be negative whilst the second is positive. This leads to a potential of the form~(\ref{Vsfs}).
An explicit construction is under investigation.

\section{Conclusion}

Reconciling moduli stabilisation and de Sitter vacua is a key issue	in the quest of an effective potential 
	appropriate  for cosmological inflation. In the type IIB string theory framework of the present work, it is shown 
	that a (suitable) non-vanishing  potential can be generated	with the internal volume modulus playing the role 
	of the inflaton $\phi$, triggering exponential growth of the Universe. 
	 Intersecting space-filling D7-branes on the other hand, are the cornerstones of the stabilisation mechanism 
	 for  K\"ahler moduli, and the creation of a positive cosmological constant in agreement with Universe's accelerated 
	 expansion today. Then, K\"ahler moduli stabilisation is achieved thanks to the logarithmic radiative corrections 
	 induced when effectively massless closed strings traverse their codimension-two bulk towards localised gravity sources. Moreover, 
	 the dS vacuum is obtained due to the positive D-term contributions  whose origin comes from $U(1)$ factors related to  
	 the intersecting D7-branes. 
	 
	 In  the large volume limit, the induced effective potential for K\"ahler  moduli  receives a  minimalist structure 
	 where its shape and in particular the volume separation, $\Delta\phi$, of its two local extrema can be parametrised in 
	 terms of a single non-negative parameter, $x$. The largest, albeit rather small $\Delta\phi$  separation occurs at a critical 
	 value $x_c>0$ where beyond this point only AdS solutions are admissible. As $x$ attains smaller values,  the distance between 
	 the two extrema diminishes and at the final admissible point $x=0$, it collapses to zero too. The upshot of the above picture 
	 is that  there exists a non-zero value $x<x_c$ at which a new inflationary small-field scenario is successfully implemented. 
	 
	 It is shown that this novel scenario is quite distinct from other well known solutions, such as hilltop inflation. 
	 Its main  benchmarks are:  
\begin{enumerate}
\item 
Most of the required number of  e-folds  ($\sim 60 $) are collected in the vicinity of   
	 the minimum of the potential, while the horizon exit arises near (from above) the inflection point. 
\item 
The corresponding inflaton field displacement is of order $10^{-2}$ compatible with the validity of the effective field theory in small-field inflation.
\item
The prediction for the tensor-to-scalar ratio of primordial density fluctuations in the early universe  is 
	 $r\approx 4\times 	10^{-4}$.  
\item  As explained above, the potential is induced by  radiative corrections which yield  a false 
	 vacuum  expected to decay to the true one towards the direction of large $\phi$ values. Implementing well established methods for  
	 the possible tunnelling~\cite{Coleman:1980aw} or its passing over the potential barrier~\cite{Hawking:1981fz}, a
	 detailed study of  the decay rate is performed and found that the false vacuum has an extremely long lifetime.  
\end{enumerate}
While inflation 
	 is successfully described close in on a sufficiently long-lived minimum of the potential, yet the cosmological constant acquires a 
	 rather large value compared to that observed today.  We describe how this problem can be evaded within the context of hybrid inflation 
	 which can be realised when a second  field  creates a new ``waterfall'' direction in the potential and inflation stops as soon as
	 the slow-roll parameter $\epsilon$ exceeds unity.  
	  
	 In closing, it is worth emphasising that the successful implementation of the cosmological inflation in the above analysis is based 
	 only on a few simple  characteristics occurring in  generic  type IIB string vacua. The few coefficients involved~\cite{Antoniadis:2019rkh}
	 depend on well defined topological properties such as the Euler characteristic of the compactification manifold, and the coefficients
	 of the D-terms determined by the geometric configuration of the intersecting D7-brane stacks.  The robustness of the results is further corroborated 
	 by the fact that, for all the measurable inflationary observables, the implications of  the logarithmic corrections and the D-terms is
	  conveyed just through  the parameter $x$ which depends only on the ratio of their coefficients and the fluxed superpotential. Consequently, 
	  the present analysis  can in principle apply to an ample class of vacua  in the string landscape.

\section *{Acknowledgements}


Work supported in part by the ERASMUS+ program and in part by a CNRS-PICS grant 07964. 
The work of GKL was supported by the ``Hellenic Foundation for
Research and Innovation (H.F.R.I.) under the `First Call for
H.F.R.I. Research Projects to support Faculty members and
Researchers and the procurement of high-cost research equipment
grant' (Project Number: 2251).''
GKL would like to thank G. Lazarides for discussions.


\end{document}